\documentclass[aip]{revtex4-2}

\usepackage{amsmath} 
\usepackage{xcolor}
\usepackage{soul}
\usepackage{physics}
\usepackage{amsfonts} 
\usepackage{chemfig} 
\setchemfig{bond style={dotted}}
\usepackage{graphicx} 
\usepackage{booktabs} 
\usepackage{subcaption}
\usepackage[version=4]{mhchem}

\draft 

\begin{document}


\title{Controlling $\langle \hat{S}^2 \rangle$ in Broken-symmetry Density
	Functional Theory  Calculations via Constrained Optimization} 



\author{Jer\'onimo Lira} \email[]{lirap1j@cmich.edu}
\affiliation{Department of Physics, Central
	Michigan University, Mount Pleasant, Michigan 48859, USA} \author{Juan E. Peralta}
\email[]{peral1j@cmich.edu}
\affiliation{Department of Physics, Central
	Michigan University, Mount Pleasant, Michigan 48859, USA}

\begin{abstract}
	Accurate determination of magnetic exchange coupling constants ($J$) from
	density functional theory (DFT) remains challenging, particularly for
	open-shell systems where broken-symmetry (BS) solutions suffer from spurious
	spin contamination that systematically exaggerates $J$ values. Several methods
	have been proposed to address this problem by adjusting the mapping scheme
	from the DFT energies to the Heisenberg-Dirac-van Vleck effective spin
	Hamiltonian energies. In this work, we explore a different route by imposing
	a constraint to the DFT energy that enforces a target value of the
	spin-squared expectation, $\langle \hat{S}^2 \rangle$, using a Lagrange
	multiplier approach. By explicitly controlling the spin character of the
	electronic state, the method attempts to overcome  limitations of standard BS
	calculations to describe magnetic interactions. As part of the theoretical
	formulation, we derive analytical expressions for the gradient of the
	spin-squared expectation value with respect to the spin-resolved density
	matrices, which are required for the practical implementation of the
	constraint within a generalized Kohn-Sham scheme. These expressions are
	general to any single-determinant method and remain valid for arbitrary spin
	states. We apply the spin-constrained approach to the calculation of $J$
	couplings and compare with  three energy-difference-based schemes for a set of
	representative systems, including \ce{H2He}, \ce{H3He3} arranged in an
	equilateral triangle, and a bis($\mu$-hydroxo) Cu(II) complex. Across all
	cases, the constrained formulation yields systematically lower and more
	consistent exchange couplings across different density functional
	approximations. This work establishes a robust and general route for
	incorporating spin-state constraints into DFT-based studies of magnetic
	exchange interactions.

\end{abstract}


\maketitle 

\section{Introduction}

Calculations based on density functional theory (DFT) have become indispensable
in electronic-structure, providing a pragmatic balance between accuracy and
computational cost,  enabling the investigation of large systems that are often
prohibitively expensive to treat with high-level post-Hartree–Fock methods, such
as coupled-cluster theory or configuration interaction.
The Hohenberg–Kohn theorems provide a rigorous theoretical foundation for DFT by
establishing that the ground-state energy is a
functional of the electron density~\cite{Hohenberg1964}. The Kohn-Sham (KS)
formalism offers a computationally tractable realization of this framework by
introducing an auxiliary non-interacting reference system that reproduces the
exact ground-state density of the interacting system~\cite{Kohn1965}.
Collectively, these  seminal developments have enabled quantitatively reliable
predictions for a broad spectrum of molecular and condensed-phase systems at a
fraction of the computational cost associated with wave-function-based
approaches. Commonly employed approximate functionals are subject to several
well-documented deficiencies that compromise reliability in specific situations.
The approximate nature of exchange–correlation functionals can give rise to
self-interaction errors~\cite{Trepte2021, Ruiz2005}, erroneous localization or
delocalization of charge~\cite{Mori2008, Bryenton2023}, and incorrect spin-state
energetics~\cite{sim2022improving, gaggioli2019beyond}. Furthermore,
conventional DFT is intrinsically formulated as a ground-state theory, rendering
it poorly suited for a faithful description of electronically excited states or
for imposing specific electronic configurations without additional
methodological extensions or constraints~\cite{herbert2023, wu2006, adamo2013,
	stein2009reliable, gould2021ensemble}. In systems whose electronic structure
substantially deviates from the variationally preferred KS solution (such as
those involving long-range charge transfer, localized magnetic moments, or
pronounced symmetry breaking) approximate functionals  may fail to reproduce the
correct qualitative and quantitative physical behavior. These limitations
underscore the need for theoretical frameworks that can systematically
incorporate physically motivated constraints that  explicitly target selected
electronic properties.

Constrained density functional theory (CDFT) generalizes the conventional
density functional framework by introducing explicit constraints on the electron
density, typically expressed as functionals of the density. In doing so, CDFT
enables access to electronic configurations that are not attainable within the
standard, unconstrained KS formalism.~\cite{dederichs1984ground, kaduk2012} The
most commonly used strategy for enforcing such constraints relies on the
incorporation of Lagrange multipliers into the variational procedure. This
formulation permits the imposition of physically motivated conditions, such as
fixed charges, spin moments, or orbital occupations, while remaining fully
embedded within the DFT framework.\cite{dederichs1984ground} As a consequence,
CDFT has proven particularly useful for the description of charge-transfer
states,\cite{wu2006direct} localized excitations,\cite{ni2025visible} and
magnetic coupling mechanisms,\cite{phillips2011magnetic, rudra2006accurate}
among a variety of other phenomena.\cite{fonseca1998self, eriksson2005many,
	hourahine2010dftb+} Crucially, CDFT preserves the computational efficiency
characteristic of standard DFT while providing enhanced control over the
electronic degrees of freedom. This additional flexibility broadens the range
of systems and properties that can be described with quantitative accuracy.

One particular context in which the limitations of standard DFT become apparent
is the broken-symmetry (BS) approach, which is extensively employed to model
open-shell singlet states and magnetic coupling in systems exhibiting spin
polarization.~\cite{ruiz1999broken, fitzhugh2023comparative} In BS DFT, the KS reference system is
allowed  to break spin symmetry in order to emulate a multi-configurational
character within a single-determinant framework. Although this strategy enables
access to low-spin (LS) solutions, the resulting states are not necessarily
eigenfunctions of the total spin operator~\cite{dema2021electronic} and in general  exhibit spin
contamination, i.e., an undesired admixture of states with different total spin
quantum numbers that complicates the physical interpretation of the computed
results.~\cite{ferre2015spin, david2020consistent} In this setting, the
expectation value of the total spin-squared operator, \(\langle \hat{S}^2
\rangle\), is routinely used as a central diagnostic to quantify the extent of
spin contamination in a given BS solution.~\cite{tsuchimochi2011constrained,
	ferre2015spin, david2020consistent} Consequently, achieving accurate control
over, or applying reliable corrections to, \(\langle \hat{S}^2 \rangle\) is
essential for a meaningful analysis of spin-dependent properties. This need has
motivated the application of constrained CDFT and related methodologies to
mitigate spin contamination and, consequently, improve the fidelity of
spin-state descriptions within a DFT framework.~\cite{andrews1991spin,
	schmidt2008controlling, ferre2015spin}

The BS approach is particularly useful for the determination of magnetic
exchange coupling constants (\(J\)), which quantify the magnetic interactions
between localized spin centers in polynuclear transition-metal complexes. These
\(J\) couplings are commonly obtained from the energy differences between HS and
BS solutions via the Heisenberg-Dirac-van Vleck (HDvV) Hamiltonian.
Consequently, the accuracy of the BS state energy has a direct and critical
influence on the predicted magnetic behavior.  Because the energy separation
between different spin states is typically small, even modest spin contamination
in the BS solution (evidenced by deviations of \(\langle \hat{S}^2 \rangle\)
from its ideal symmetry-broken reference value) can introduce substantial errors
into the calculated \(J\) constants. This spin contamination complicates the
reliable approximation of BS solutions, motivating the development of various
spin-decontamination schemes to address this problem.~\cite{andrews1991spin,
	ovchinnikov1996simple, ferre2015spin}
Alternative strategies include reformulating the DFT-to-Heisenberg mapping~\cite{Moreira2006}
or using time-dependent DFT to access proper spin eigenstates.~\cite{rudra2006accurate}
An early related approach was proposed by L\"owdin, which consists in 
projecting the broken-symmetry determinant onto pure spin eigenstates using
group-theoretical projection operators.~\cite{lowdin1964angular} While Löwdin's method guarantees 
that at least one projected state lies lower in energy than the original 
determinant, it requires numerical integration over the rotation group and
becomes computationally demanding for large systems.

In addition, standard BS-DFT calculations are widely reported to overestimate
the magnitude of \(J\), particularly when using
local and semi-local density functional approximations~\cite{rivero2008reliability},
further motivating approaches that enforce a more controlled spin character in 
the underlying BS reference.
Thus, explicit control of \(\langle \hat{S}^2 \rangle\) via CDFT methodologies 
becomes essential for obtaining dependable \(J\) values and for ensuring that 
the resulting magnetic couplings reflect the genuine electronic structure rather
than artifacts arising from spin-state mixing.
This work introduces  such an approach within the
generalized KS (GKS) formalism, wherein \(\langle \hat{S}^2 \rangle\) is
explicitly constrained during the self-consistent field (SCF) procedure via a
Lagrange multiplier.
By guiding the optimization of the spin-resolved density matrices toward target
spin values, this method drives BS states toward the desired
\(\langle \hat{S}^2\rangle\) and, to a lesser extent, reduces residual spin
contamination in HS states.

It is important to remark that some BS solutions are intrinsically affected by
spin contamination, and the aim of the present approach is not to remove this
contamination entirely, but rather to control it in a physically motivated  way.
In the symmetry-broken limit~\cite{perdew2025scan}, BS states correspond to
well-defined linear combinations of spin eigenstates, which in turn yield
characteristic target values of \(\langle \hat{S}^2 \rangle\). In practical DFT
calculations, however, these ideal values are seldom reproduced exactly. The
deviations originate from the approximate nature of the exchange-correlation
functionals, and in particular from self-interaction error, which favors
excessive orbital delocalization and consequently alters the relative weights of
the spin components.~\cite{mori2006many} As an illustrative  case, consider the
linear H–He–H molecule. The BS solution representing the open-shell singlet
state corresponds, in the symmetry-broken limit, to a 50:50 admixture of singlet
(\(S = 0\), \(\langle \hat{S}^2 \rangle = 0\)) and triplet (\(S = 1\), \(\langle
\hat{S}^2 \rangle = 2\)) components, which leads to \(\langle \hat{S}^2 \rangle
= 1\). In practice, one typically obtains values slightly below this ideal limit
(e.g., \(\langle \hat{S}^2 \rangle \approx 0.97\)). Although such discrepancies
may appear numerically small, they can induce non-negligible variations in the
total energy. Since the evaluation of the exchange coupling constant \(J\) is
highly sensitive to the energy difference between the HS and BS solutions, even
minor deviations from the target \(\langle \hat{S}^2 \rangle\) value can lead to
substantial changes in the resulting \(J\) parameters. The capability to impose
a consistent target \(\langle \hat{S}^2 \rangle\) therefore offers a controlled
and systematic strategy to mitigate these uncertainties.

The remainder of this article is organized as follows. Section~\ref{sec:theory}
presents the theoretical framework and computational details of our CDFT
implementation, including the variational formulation with a spin-squared
constraint, analytical derivatives, and the exchange-coupling schemes used in
this work. In Section~\ref{sec:results}, we assess the performance of the method
for \ce{H2He}, \ce{H3He3}, and the bis($\mu$-hydroxo) Cu(II) complex by
comparing \(J\) values from unconstrained and constrained energies.
Section~\ref{sec:conclusions} summarizes our findings.

\section{Theory and Computational Details}\label{sec:theory}

The ground-state DFT  energy is obtained by minimizing an energy functional
expressed in terms of the electron spin densities \(n^\sigma\), where
\(\sigma=\alpha, \beta\) denotes the two possible spin projections in the
collinear spin-polarization framework.~\cite{von1972local, perdew1981,
	chattaraj2009chemical} The KS energy functional is written as:
\begin{equation}
	\label{KSfunctional}
	E_\text{DFT}[n^\alpha, n^\beta] = \sum_a \left\langle
	\psi_a \left| -\frac{1}{2}\nabla^2 \right| \psi_a \right\rangle + E_H[n] +
	\int v(\mathbf{r}) n(\mathbf{r})\,d^3\mathbf{r} + E_\text{XC}[n^\alpha, n^\beta]
	\; ,
\end{equation} where \(n = n^\alpha + n^\beta\) is the total electron
density, the first term in the  r.h.s. of Eq~\ref{KSfunctional} represents the
kinetic energy of non-interacting electrons, \( E_H[n]\) is the classical
Coulomb energy (also known as Hartree energy), \( v(\mathbf{r}) \) is the
external potential including the electron-nuclei interaction, and \(
E_\text{XC}[n^\alpha, n^\beta] \) is the exchange-correlation energy functional,
which accounts for all the quantum mechanical effects beyond the classical
electrostatics.

The constraint problem can be formulated through the Lagrangian formulation,
\begin{equation}
	\label{Lagrangian} W[n^\alpha, n^\beta, \lambda] =
	E_\text{DFT}[n^\alpha, n^\beta] + \lambda  \left( \sum_\sigma\int
	w^\sigma(\mathbf{r}) n^\sigma(\mathbf{r})\,d^3\mathbf{r} - C \right) \; ,
\end{equation}
where \(\lambda\) is the Lagrange multiplier, \(w^\sigma(\mathbf r)\) are
prescribed weight functions defining the constrained quantity, and \(C\) is its
target value.
This Lagrangian is made stationary with respect to the variational variables
\(n^\sigma(\mathbf r)\) and \(\lambda\).
Stationarity with respect to the Lagrange multiplier,
\(\partial W/\partial\lambda=0\), enforces the constraint,
\begin{equation}
	\sum_\sigma\int w^\sigma(\mathbf r)\,n^\sigma(\mathbf r)\,d^3\mathbf r = C \; .
\end{equation}
The choice of the constrained target quantity \(C\) depends on the physical
property we aim to control.

In the case of constraining \(\langle \hat{S}^2 \rangle\), the corresponding
Lagrangian  can be cast as
\begin{equation}
	\label{Lagrangian2}
	W[n^\alpha, n^\beta, \lambda] = E_\text{DFT}[n^\alpha, n^\beta] + \lambda \left(\langle \hat{S}^2 \rangle -
	S_c^2\right) \,,
\end{equation}
where \(S_c^2\) is the spin-square target value.
Although CDFT typically requires the
constrained quantities to be expressed as explicit functionals of
\(n^\sigma\), as in Eq.~\ref{Lagrangian},
this is not possible in practice for two-electron operators such as
\(\langle \hat{S}^2 \rangle\) since the   functional form \(\langle \hat{S}^2
\rangle[n^\sigma]\) is unknown. A formal caveat is that 
\(\langle \hat{S}^2 \rangle\) is not rigorously defined as an exact density 
functional in DFT since  the Kohn--Sham determinant is only an auxiliary object 
used to construct the density. As a result, in practical DFT calculations,
\(\langle \hat{S}^2 \rangle\) is evaluated using the spin-resolved one-particle
reduced density matrices (1-RDM). 
Expressed in an atomic orbital basis, the
matrix element of the 1-RDM can be written as
\begin{equation}
	P_{\mu\nu}^{\sigma} = \sum_{i \in \sigma} f^\sigma_i \, C^\sigma_{\mu
		i} C_{\nu i}^{\sigma*} \; ,
\end{equation}
where $C^\sigma_{\nu i}$ are the molecular orbital coefficients, and $f^\sigma_i$ the occupations.

While \(n^\sigma\) can be evaluated from the spin-resolved density matrices
$\mathbf{P^\sigma}$ and the basis functions, the inverse mapping, from
$n^\sigma$ to a unique $\mathbf{P^\sigma}$, is not well-defined. This results in
the unknown form of the \(\delta\langle \hat{S}^2 \rangle/\delta n^\sigma\)
functional derivative in the   the Lagrangian minimization. However, in GKS, the
minimization is performed varying the 1-RDM, so we can circumvent this problem
using  the derivative of \(\langle \hat{S}^2 \rangle\) with respect to the
1-RDM, which is well-defined in the finite basis used and readily accessible
throughout the SCF cycle.
With this, the average spin-squared value can be incorporated into the
constrained DFT framework by applying the constraint directly on \(\langle
\hat{S}^2[\mathbf{P}^\alpha, \mathbf{P}^\beta]\rangle\), enabling the variational minimization of
the GKS energy.~\cite{andrews1991spin, tsuchimochi2011constrained}

To minimize the  Lagrangian for a fixed value of
$\lambda$, it is convenient to define an effective GKS Hamiltonian that
incorporates the constrain as the derivative of the Lagrangian  with respect to
\(\mathbf{P^\sigma}\),~\cite{lehtola2020overview}
\begin{align}
	F_{\mu\nu}^{\text{c},
	\sigma} & = \frac{d}{d P^\sigma_{\mu\nu}} W[n^\alpha, n^\beta, \lambda] \\ &= F^\sigma_{\mu\nu} +
	\lambda \frac{d}{d P^\sigma_{\mu\nu}}\langle \hat{S}^2 \rangle \, ,
\end{align}
with these effective GKS Hamiltonian matrices $\mathbf{F^\sigma}$ replacing the
standard KS Hamiltonian. In order to evaluate the contribution of the constraint
to the $\mathbf{F^\sigma}$  matrices, it is necessary to express \(\langle
\hat{S}^2 \rangle\) explicitly in terms of the spin-resolved density matrices,
\begin{equation}
	\label{eq:s2}
	\langle \hat{S}^2 \rangle =
	S_z(S_z + 1) + N_\beta -
	\sum_{\mu\nu\kappa\lambda}P^\alpha_{\mu\nu}O_{\nu\kappa}P^\beta_{\kappa\lambda}O_{\lambda\mu} \; ,
\end{equation}
where \(S_z = (N_\alpha - N_\beta)/2\), \(N_\sigma = \sum_\mu
P^\sigma_{\mu\nu}O_{\nu\mu}\), and \(O\) is the standard atomic orbital overlap
matrix. This expression is general for any number of unpaired electrons in
monodeterminantal methods such as Hartree Fock and DFT. 
The spin-contamination contribution vanishes when
\(N_\beta = \sum_{\mu\nu\kappa\lambda}P^\alpha_{\mu\nu}O_{\nu\kappa}P^\beta_{\kappa\lambda}O_{\lambda\mu}\);
a convenient set of conditions under which this equality holds is discussed and
proved in Appendix~\ref{appendixb}.
We introduce the derivatives of Eq.~\ref{eq:s2}:
\begin{equation}
	\label{eq:s2grada}
	\frac{\partial{\langle \hat{S}^2 \rangle}}{\partial P^\alpha_{\mu\nu}} =
	\frac{1}{2}(N_\alpha - N_\beta + 1) O_{\nu\mu} - [\mathbf{O} \mathbf{P}^\beta \mathbf{O}]_{\nu\mu} \; ,
\end{equation}
and
\begin{equation}
	\label{eq:s2gradb}
	\frac{\partial{\langle \hat{S}^2 \rangle}}{\partial P^\beta_{\mu\nu}} =
	\frac{1}{2}(N_\beta - N_\alpha + 1) O_{\nu\mu} - [\mathbf{O} \mathbf{P}^\alpha \mathbf{O}]_{\mu\nu} \; .
\end{equation}
We note that  previous constrain approaches focused  on suppressing the
spin-contamination term, \(N_\beta - \sum_{\mu\nu\kappa\lambda}
P^\alpha_{\mu\nu} O_{\nu\kappa} P^\beta_{\kappa\lambda} O_{\lambda\mu}\), in
Eq.~\ref{eq:s2}. Such approach does not allow direct access to BS states with a
prescribed total spin.~\cite{schmidt2008controlling, andrews1991spin} In
contrast, by constraining the full expectation value \(\langle \hat{S}^2
\rangle\), our approach enables us, for example,  to explicit target the BS
singlet state with an arbitrary \(\langle \hat{S}^2 \rangle \) value.

Finally, the constrained solution corresponds to a stationary point of the
Lagrangian \(W[n^\alpha, n^\beta,\lambda]\) with respect to both the
spin-resolved density matrices \(\mathbf{P}^\sigma\) and the Lagrange multiplier
$\lambda$. In practice, the electronic structure is optimized self-consistently
using the modified GKS matrices, while $\lambda$ is adjusted (in an outer loop)
until the target condition \(\langle \hat{S}^2 \rangle = S_c^2\) is satisfied;
the specific optimization strategy and convergence thresholds are given at the
end of this section.

Having established the self-consistent procedure for enforcing a target
\(\langle \hat{S}^2 \rangle\), we now turn to its application in the context of
magnetic interactions. In particular, we use the constrained formalism to
compute isotropic magnetic exchange coupling \(J\), which for the case two centers $A$
and $B$ is defined though the  effective Heisenberg spin Hamiltonian~\cite{caballol1997remarks, postnikov2006density},
\begin{equation}
	\hat{H} = -2J\, \hat{\mathbf{S}}_A \cdot \hat{\mathbf{S}}_B \; ,
\end{equation}
where \(\hat{\mathbf{S}}_A\) and $\hat{\mathbf{S}}_B$ are spin operators. For
comparison, we  compute the exchange coupling constant \(J\) using three
established schemes commonly employed in BS DFT. These expressions estimate
\(J\) from the energy difference between a LS state and a HS state, where LS
configuration is approximated by a BS solution and the HS configuration by a
spin-pure HS reference. Each scheme relies on distinct assumptions regarding the
spin character of these states and their mapping onto the effective Heisenberg
Hamiltonian.

We first apply our CDFT approach using Noodleman's
expression~\cite{noodleman1981valence},
\begin{equation}\label{eq:noodleman}
	J_N = \frac{E^{\text{BS}} - E^{\text{HS}}}{S^2_{\text{max}}} \; .
\end{equation}
This expression is derived from a spin-projection argument and is formally valid
in the weak-overlap limit between the magnetic orbitals. It assumes that the BS
solution can be approximated as an equal-weight linear combination of the HS and
LS eigenstates. Within this framework, the energy difference is rescaled by the
maximum spin eigenvalue \(S_{\text{max}}\) to recover an effective exchange
coupling constant. In the present work, the target 
\(\langle \hat{S}^2 \rangle\) values used for the BS solutions are chosen to 
match the spin mixing assumed in Noodleman's BS-to-Heisenberg mapping. A 
necessary clarification is that the spin-squared constraint
is also applied to the nominally spin-pure HS state, as practical DFT
calculations generally exhibit small but nonzero deviations from the ideal
\(\langle \hat{S}^2 \rangle\) value.

In addition to Noodleman's formulation, we also evaluate the exchange coupling
constant using two alternative schemes commonly employed in BS DFT. The second
scheme, due to Ruiz and co-workers~\cite{ruiz1999broken}, is a non-projected
approach designed for the strong coupling limit:
\begin{equation}\label{eq:ruiz} J_R = \frac{E^{\text{BS}} -
		E^{\text{HS}}}{S_{\text{max}}(S_{\text{max}} + 1)} \; .
\end{equation}
Here, the BS solution is treated as a good approximation to the LS state itself,
without attempting to correct for spin contamination. This approach is widely
used in the literature, especially when the BS and LS states are close in
character.

The third scheme is Yamaguchi's approach,~\cite{soda2000ab} also known as
generalized spin-projected, which introduces a more flexible, generalized
spin-projected expression:
\begin{equation}\label{eq:yamaguchi}
	J_Y = \frac{E^{\text{BS}} - E^{\text{HS}}}{\langle \hat{S}^2
		\rangle^{\text{HS}} - \langle \hat{S}^2 \rangle^{\text{BS}}} \; .
\end{equation}
This method accounts for spin contamination by using the expectation
values of the spin-squared operator in the mapping scheme.
It interpolates between the weak and strong coupling regimes, making it
particularly suited for systems where the degree of spin contamination
varies significantly.
In the present work,  only Noodleman's  approach is considered in combination
with the   $\langle \hat{S}^2 \rangle$ constraint, and the remaining approaches
are used with unconstrained DFT  solely for comparison. This restriction is
motivated by conceptual consistency: Ruiz's scheme assumes strong coupling where
the BS state approximates the LS state directly, without the equal-weight mixing
assumed in Noodleman's formalism. For Yamaguchi's scheme, applying the
constraint would be redundant since enforcing the target $\langle \hat{S}^2
	\rangle$ values causes its denominator to reduce to $S_{\text{max}}^2$,
recovering Noodleman's expression. Thus, these unconstrained schemes serve as
benchmarks for assessing how standard DFT deviations from ideal spin mixing
affect $J$ across different theoretical frameworks.

Calculations were performed using the Python-based PySCF
package~\cite{pyscf2020, LEHTOLA20181}. Two model systems were considered
initially: the linear triatomic molecule \ce{H2He} and the equilateral
\ce{H3He3} cluster. For both systems, we employed the 6-311G** basis
set.~\cite{krishnan1980self} To assess the method on a more realistic system, we
also studied the bis($\mu$-hydroxo) Cu(II) complex. For this case, atomic
coordinates were taken from Ref.~\citenum{singh2021}, and for comparison with
previous studies the def2-TZVP basis set~\cite{weigend2005} was used. All calculations were carried
out within the unrestricted KS (UKS) framework. For the \ce{H2He} and \ce{H3He3}
systems, we employed the PBE~\cite{perdew1996generalized},
BLYP~\cite{becke1988density}, PBEh~\cite{adamo1999toward},
B3LYP~\cite{becke1993density} and SCAN~\cite{sun2015strongly} approximate
exchange-correlation functionals, representing generalized gradient
approximation (GGA), hybrid, and meta-generalized gradient approximation
(meta-GGA) families, respectively. For each system and functional, both the HS
and BS solutions were obtained. The HS state was generated by initializing all
unpaired electrons with parallel spins. The BS solution was then obtained by
performing a single spin flip on one of the unpaired electrons to generate an
initial guess and allowing the calculation to converge to the corresponding
antiferromagnetic BS configuration.

The Lagrange multiplier \(\lambda\) was optimized using the
Broyden-Fletcher-Goldfarb-Shanno (BFGS) quasi-Newton
method~\cite{nocedal2006numerical} as implemented in
\texttt{scipy.optimize.minimize}, with analytical gradients supplied to
accelerate convergence (Eq.~\ref{eq:s2grada} and Eq.~\ref{eq:s2gradb}). 
The SCF convergence thresholds were set to \(10^{-10}\) Hartree for the total energy.
After SCF convergence at fixed \(\lambda\), the value of \(\langle \hat{S}^2 \rangle\) was
evaluated and compared to the target \(S_c^2\). The Lagrange multiplier was then
updated using the BFGS optimizer until the constraint condition was satisfied,
with the absolute deviation \(|\langle \hat{S}^2 \rangle - S_c^2|\) below
\(10^{-5}\). This outer-loop structure allows the electronic degrees of freedom
and the Lagrange multiplier to be optimized in a decoupled and numerically
stable manner.
A more detailed discussion of the dependence of the total energy and
\(\langle \hat{S}^2 \rangle\) on \(\lambda\), and of the
qualitatively different behavior observed for HS and BS states, is provided in
Appendix~\ref{appendixa}.

\section{Results}\label{sec:results}

In this section, we present \(J\) couplings obtained from both unconstrained GKS
calculations and the spin-constrained CDFT approach introduced in
Section~\ref{sec:theory}. The unconstrained results are analyzed using the three
schemes described in the previous Section, and  given by
Eqs.~\ref{eq:noodleman}, \ref{eq:ruiz}, and \ref{eq:yamaguchi}, respectively.
These values are compared with the exchange coupling constant obtained from the
constrained approach, \(J_c\), which is evaluated using Noodleman's expression,
Eq.~\ref{eq:noodleman}, with spin-constrained energies for both the HS and LS
states.

This systematic comparison enables an assessment of how the constraint
influences the magnitude, functional dependence, and physical consistency of the
computed magnetic couplings across different couplings schemes.
\begin{figure}[ht]
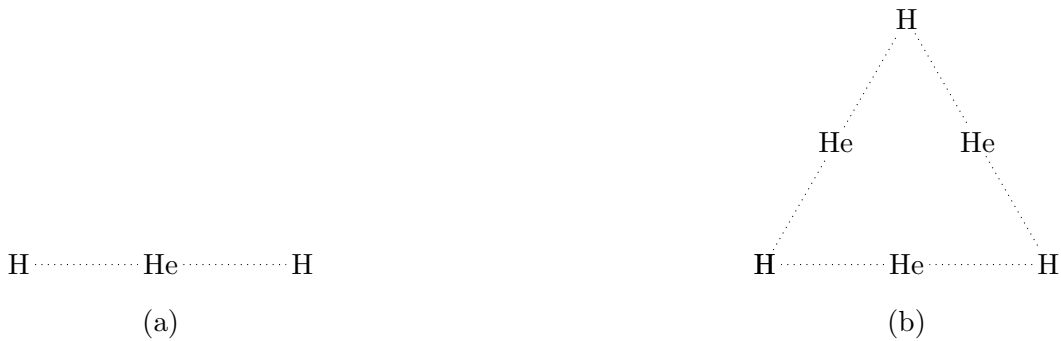

	\centering
	\begin{subfigure}[b]{0.4\textwidth}
		\centering
		\chemfig{H-[:0,1.625]He-[:0,1.625]H}
		\caption{}
	\end{subfigure}
	\hfill
	\begin{subfigure}[b]{0.4\textwidth}
		\centering
		\chemfig{
		H
		-[:60,1.625]He
		-[:60,1.625]H
		-[:-60,1.625]He -[:-60,1.625]H -[:180,1.625]He -[:180,1.625]H }
		\caption{}
	\end{subfigure}
	\caption{Schematic representations of the two prototypical models
		considered: the linear \ce{H2He} molecule (left), for which the He--H bond
		distance is systematically varied (1.25, 1.625, and 2.00~\AA), and the
		equilateral \ce{H3He3} cluster (right), with a fixed side length of
		3.25~\AA.}
	\label{fig:geometries}
\end{figure}

\subsubsection*{\ce{H2He}}

We first consider the linear \ce{H2He} model system, in which the two terminal
hydrogen atoms act as magnetic centers coupled through a closed-shell helium
bridge, Fig.~\ref{fig:geometries}(a). To probe the distance dependence of the
exchange interaction, three symmetric geometries were studied with He--H bond
lengths of 1.25, 1.625, and 2.00~\AA. The HS reference closely corresponds to
the triplet state (\(S=1\)), whereas the LS configuration is a BS singlet
(\(S=0\)), which may exhibit spin contamination in unconstrained calculations
and is therefore the main target of the \(\langle \hat{S}^2\rangle\) constraint.

Tables~\ref{HHeH125}–\ref{HHeH200} report \(J_N\), \(J_R\), \(J_Y\) and \(J_c\)
(in cm\(^{-1}\)) for the \ce{H2He} molecule at three different He–H bond
distances. At the shortest distance (1.25 \AA), the system lies in the strong
coupling regime, where the magnetic centers interact through significant orbital
overlap. In this case, the unconstrained BS states exhibit the largest deviation
of \(\langle \hat{S}^2 \rangle\) with respect to the ideal value of 1, as shown
in the third column of Table~\ref{HHeH125}. As a consequence, enforcing the
constraint on \(\langle \hat{S}^2 \rangle\) leads to substantial reductions in
the computed \(J\). The effect is especially evident relative to the
conventional DFT schemes \(J_N\) and \(J_Y\), for which the spin-projection
factor enters explicitly through the denominator of Eqs.~\ref{eq:noodleman} and
\ref{eq:yamaguchi}.

Ruiz's formula  performs better in this regime, since the assumptions underlying
its formulation explicitly account for a strong overlap between the
magnetic-center orbitals. It is important to remark that unconstrained DFT
systematically overestimates the exchange coupling strength in this regime.
Although the constrained values \(J_c\) are considerably smaller  than their
unconstrained counterparts, they remain significantly smaller than and
substantially depart from the values obtained from Full-CI.
\begin{table}[ht]
	\begin{ruledtabular}
		\caption{\(J_N\), \(J_R\), and \(J_Y\) (in cm\(^{-1}\)) for the \ce{H2He} system at a He--H bond distance of 1.25~\AA.}
		\label{HHeH125}
		\begin{tabular}{lccccccr}
			Functional & \(\langle\hat{S}^2\rangle_\text{HS}\) & \(\langle\hat{S}^2\rangle_\text{BS}\) & \(J_N\) & \(J_R\) & \(J_Y\) & \(J_c\) & Full-CI\footnote{Taken from Ref.~\citenum{soda2000ab}} \\
			\hline
			PBE        & 2.00094                               & 0.68264                               & -4567   & -2283   & -3465   & -830    & -2430                                                  \\
			BLYP       & 2.00101                               & 0.58507                               & -5391   & -2695   & -3807   & -925                                                             \\
			PBEh       & 2.00098                               & 0.81892                               & -3647   & -1823   & -3085   & -792                                                             \\
			B3LYP      & 2.00105                               & 0.74330                               & -4366   & -2183   & -3471   & -863                                                             \\
			SCAN       & 2.00122                               & 0.76966                               & -4333   & -2166   & -3520   & -861                                                             \\
		\end{tabular}
	\end{ruledtabular} \end{table}
At intermediate distances (1.625 \AA), both the magnetic coupling strength and
the degree of spin contamination are reduced. In this regime, the singlet state
acquires an increasing multiconfigurational character, making the constraint
applied to the BS solution more physically justified and, therefore, central to
the objective of our approach. At this distance, the energy differences between
the unconstrained KS-DFT calculations and the constrained approach are
noticeably smaller than those observed in the short-distance regime discussed
above. Although CDFT still leads to a reduction of the computed \(J_c\), the
relative impact of the constraint is less pronounced. As shown in
Table~\ref{HHeH1625}, the \(\langle \hat{S}^2 \rangle\) values associated with
the BS state are significantly closer to the ideal value of 1. The \(\langle
\hat{S}^2 \rangle\) value for the HS state continues to exhibit only a small
degree of spin contamination and shows no significant dependence on the bond
length when compared to the short-distance case.
\begin{table}[ht]
	\begin{ruledtabular}
		\caption{\(J_N\), \(J_R\), and \(J_Y\) (in cm\(^{-1}\)) for the \ce{H2He} system at a He--H bond distance of 1.625~\AA.}
		\label{HHeH1625}
		\begin{tabular}{lccccccr}
			Functional & \(\langle\hat{S}^2\rangle_\text{HS}\) & \(\langle\hat{S}^2\rangle_\text{BS}\) & \(J_N\) & \(J_R\) & \(J_Y\) & \(J_c\) & Full-CI\footnote{Taken from Ref.~\citenum{soda2000ab}} \\
			\hline
			PBE        & 2.00035                               & 0.97727                               & -472    & -236    & -461    & -170    & -272                                                   \\
			BLYP       & 2.00043                               & 0.96355                               & -621    & -310    & -599    & -217                                                             \\
			PBEh       & 2.00035                               & 0.98488                               & -390    & -195    & -384    & -159                                                             \\
			B3LYP      & 2.00043                               & 0.97589                               & -512    & -256    & -500    & -196                                                             \\
			SCAN       & 2.00047                               & 0.97697                               & -474    & -237    & -463    & -165                                                             \\
		\end{tabular}
	\end{ruledtabular}
\end{table}
At the longest distance (2.00 \AA), Table~\ref{HHeH200}, the H atoms are only
weakly coupled through the He bridge. In this weak overlap regime, the BS
solution exhibits an increased degree of spin contamination; the \(\langle
\hat{S}^2 \rangle\) value of the BS state is even closer to 1. As a result, the
energy corrections introduced by the constraint are small. In this regime,
Yamaguchi's formula effectively reduces to  Noodleman's expression. Since the
latter assumes a maximally multiconfigurational character in the weak-coupling
limit, this behavior is naturally captured by  Yamaguchi's scheme, as reflected
by the very similar values of the \(J_N\) and \(J_Y\) in columns 4 and 6,
respectively.
\begin{table}[ht]
	\begin{ruledtabular}
		\caption{\(J_N\), \(J_R\), and \(J_Y\) (in cm\(^{-1}\)) for the \ce{H2He} system at a He--H bond distance of 2.0~\AA.}
		\label{HHeH200}
		\begin{tabular}{lccccccr}
			Functional & \(\langle\hat{S}^2\rangle_\text{HS}\) & \(\langle\hat{S}^2\rangle_\text{BS}\) & \(J_N\) & \(J_R\) & \(J_Y\) & \(J_c\) & Full-CI\footnote{Taken from Ref.~\citenum{soda2000ab}} \\
			\hline
			PBE        & 2.00008                               & 0.99824                               & -45     & -22     & -45     & -25     & -25                                                    \\
			BLYP       & 2.00014                               & 0.99675                               & -69     & -34     & -69     & -42                                                              \\
			PBEh       & 2.00008                               & 0.99878                               & -37     & -18     & -37     & -23                                                              \\
			B3LYP      & 2.00013                               & 0.99770                               & -56     & -28     & -56     & -35                                                              \\
			SCAN       & 2.00014                               & 0.99828                               & -39     & -19     & -39     & -24                                                              \\
		\end{tabular}
	\end{ruledtabular} \end{table}
\begin{figure}[ht] \centering
	\begin{subfigure}[b]{0.49\textwidth}
		\centering
		\includegraphics[width=\textwidth]{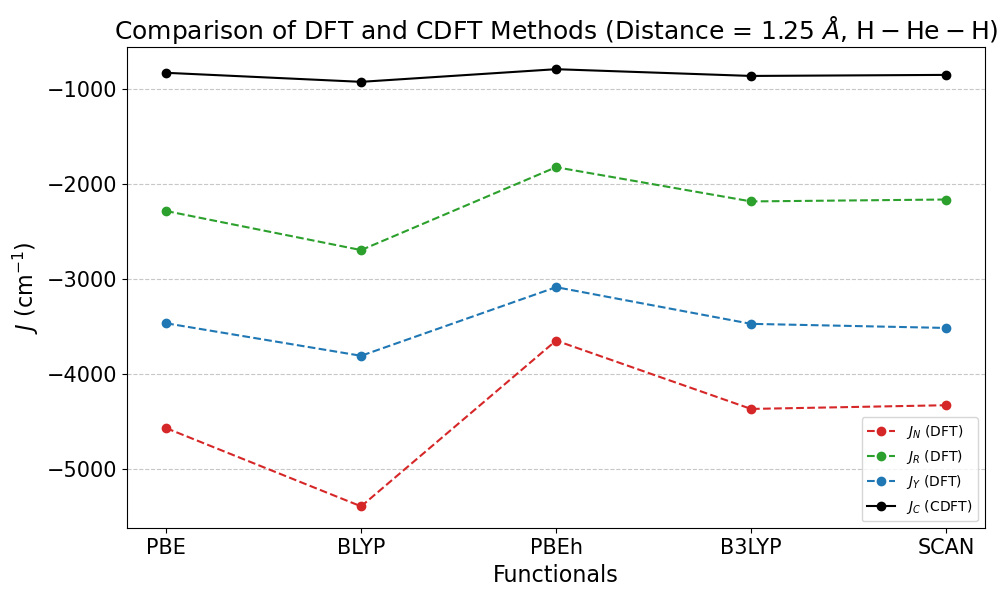}
		\caption{}
	\end{subfigure}
	\hfill
	\begin{subfigure}[b]{0.49\textwidth}
		\centering
		\includegraphics[width=\textwidth]{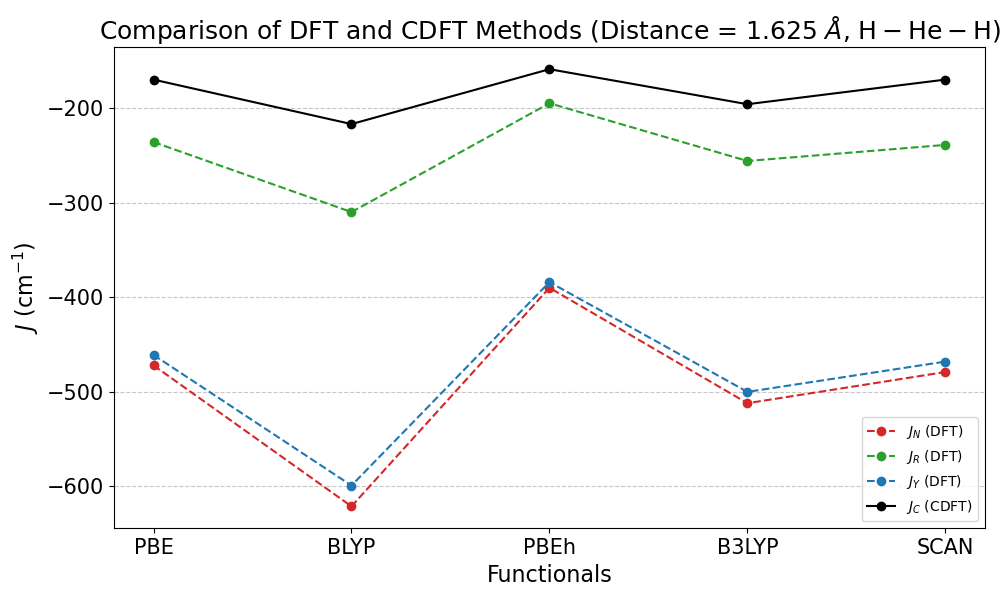}
		\caption{}
	\end{subfigure}
	\vspace{0.4cm}

	\begin{subfigure}[b]{0.49\textwidth}
		\centering
		\includegraphics[width=\textwidth]{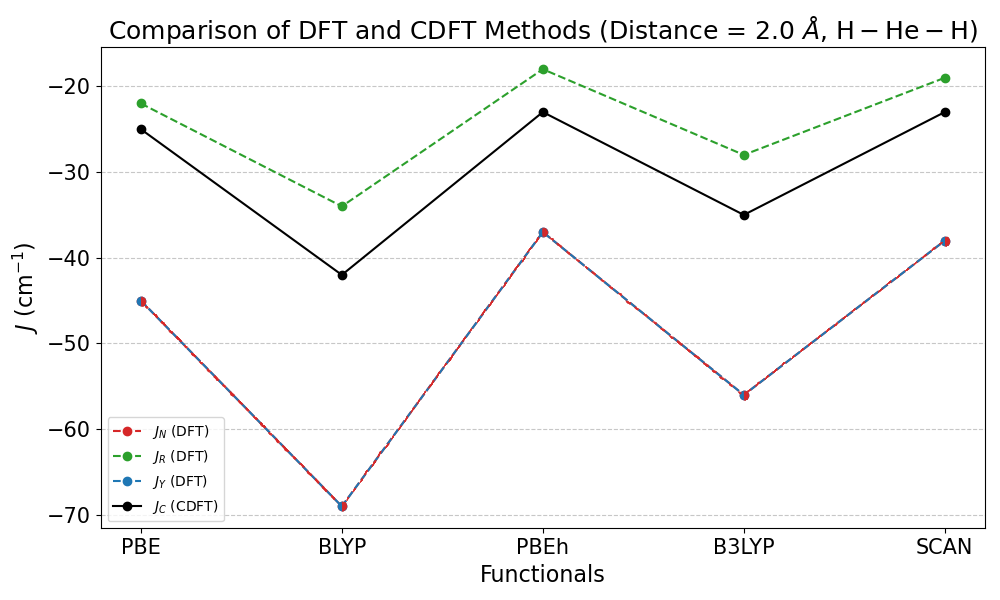}
		\caption{}
		\label{fig:hheh_3}
	\end{subfigure}
	\hfill
	\begin{subfigure}[b]{0.49\textwidth}
		\centering
		\includegraphics[width=\textwidth]{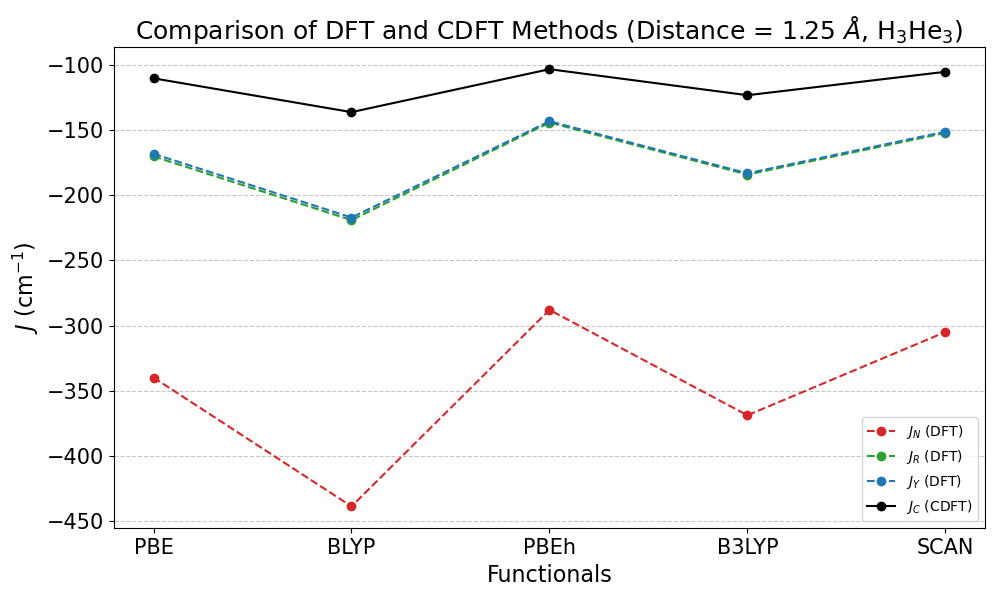}
		\caption{}
	\end{subfigure}
	\caption{Dependence of the total energy and the spin-squared expectation
		value \(\langle \hat{S}^2 \rangle\) on the Lagrange multiplier \(\lambda\) for
		the \ce{H2He} system at three different He--H bond distances. And, on the
		bottom
		right, for the triangular \ce{H3He3}}
	\label{fig:hheh_all}
\end{figure}
Figure~\ref{fig:hheh_all}, in panels a, b and c, provides a graphical
representation of the \(J\) reported in Tables~I–III for the \ce{H2He} system at
the three He--H bond distances considered. Dotted lines denote values of \(J\)
obtained from unconstrained DFT calculations, while the continuous lines
correspond to results obtained using the spin-constrained DFT formalism. These
lines connecting the different dots are included as a guide to the eye to
emphasize trends across different exchange–correlation functionals. At 1.25 and
1.625 \AA, the constrained calculations yield exchange-coupling constants that
are less sensitive to the choice of exchange–correlation functional. This
reduction in functional dependence is particularly evident when compared to the
broader dispersion observed in the unconstrained DFT results. In addition, the
constrained \(J\) values are systematically smaller in magnitude, as expected
given the commonly reported overestimation of magnetic exchange interactions by
standard BS-DFT. In Fig.~\ref{fig:hheh_3}, the values corresponding to the
Yamaguchi formula and the Noodleman formula coincide. Overall,
Fig.~\ref{fig:hheh_all} demonstrates that enforcing the \(\langle \hat{S}^2
\rangle\) constraint not only lowers the magnitude of the \(J\) but also
stabilizes their functional dependence. This increased consistency supports the
interpretation that the constrained framework yields more physically meaningful
and transferable estimates of magnetic exchange interactions than unconstrained
BS-DFT.

\subsubsection*{\ce{H3He3}}

We next consider the triangular \ce{H3He3} cluster shown in
Fig.~\ref{fig:geometries}(b), which introduces a multi-center coupling topology
absent in the linear \ce{H2He} model. The three hydrogen atoms occupy the
vertices of an equilateral triangle with side length 3.25~\AA, while the three
helium atoms are located at the midpoints of each edge, resulting in a fixed
H--He distance of 1.625~\AA. The HS reference corresponds to a quartet state
(\(S = 3/2\)), whereas the LS configuration is described by a doublet (\(S =
1/2\)). As in the linear case, the BS solution may exhibit spin contamination.
Accordingly, we apply the \(\langle \hat{S}^2 \rangle\) constraint to enforce a
target value of 1.75, corresponding to an equal admixture of a doublet and a
quartet. The triangular geometry introduces competing magnetic interactions
between the three hydrogen centers, which can give rise to spin frustration,
i.e., a situation in which not all pairwise spin couplings can be simultaneously
optimized. This geometric constraint is therefore expected to weaken the
effective magnetic coupling relative to the linear \ce{H2He} system.

Table~\ref{H3He3table} shows  exchange couplings  obtained from unconstrained
DFT calculations, and  the constrained value \(J_c\) computed within the CDFT
framework. Consistent with the behavior observed for the linear system, imposing
the spin-squared constraint systematically reduces the magnitude of the exchange
coupling constants across all functionals.
\begin{table}[ht]
	\begin{ruledtabular}
		\caption{\(J_N\), \(J_R\), and \(J_Y\) (in cm\(^{-1}\)) for the \ce{H3He3} system at a He--H bond distance of 1.625~\AA.}
		\label{H3He3table}
		\begin{tabular}{lccccccr}
			Functional & \(\langle\hat{S}^2\rangle_\text{HS}\) & \(\langle\hat{S}^2\rangle_\text{BS}\) & \(J_N\) & \(J_R\) & \(J_Y\) & \(J_c\) & CASPT2\footnote{Taken from Ref.~\citenum{Ruiz2005}} \\
			\hline
			PBE        & 3.75079                               & 1.73277                               & -340    & -170    & -168    & -110    & -195                                                \\
			BLYP       & 3.75087                               & 1.72537                               & -439    & -219    & -217    & -136                                                          \\
			PBEh       & 3.75073                               & 1.73827                               & -288    & -144    & -143    & -103                                                          \\
			B3LYP      & 3.75082                               & 1.73287                               & -369    & -184    & -183    & -123                                                          \\
			SCAN       & 3.75091                               & 1.73396
			           & -305                                  & -152                                  & -151    & -105                                                                              \\
		\end{tabular}
	\end{ruledtabular}
\end{table}
The values of \(\langle \hat{S}^2 \rangle\) obtained for the HS state are very
close to the expected value of 3.75, reflecting its nearly spin-pure character.
As a result, enforcing the constraint to recover the exact value of \(\langle
\hat{S}^2 \rangle = 3.75\) leads to only minor corrections to the HS energies,
which nevertheless contribute to the evaluation of the constrained exchange
coupling constant \(J_c\). In contrast, the BS solutions exhibit larger
deviations from the ideal target value of 1.75. With the exception of BLYP,
which yields slightly lower \(\langle \hat{S}^2 \rangle\) values, the BS spin
expectation values are similar across the different functionals. Despite this
variation, the resulting constrained exchange couplings \(J_c\) are comparable
for all functionals.

When compared to the linear \ce{H2He} system at a similar bond-lengths, the
absolute values of the \(J\) for \ce{H3He3} are smaller. This reduction is
consistent with a weakened effective magnetic coupling arising from spin
frustration and a more delocalized distribution of the spin density over the
triangular cluster.

\subsubsection*{Bis($\mu$-hydroxo) Cu(II)}

\begin{figure}[ht]
 \caption{Molecular structure of the bis($\mu$-hydroxo) Cu(II) complex with
		bridging hydroxo ligands. The Cu--Cu distance of 2.97~\AA{} and the Cu--O--Cu
		angle give rise to an angle-dependent superexchange pathway. \label{fig:cu2}
        }
        \includegraphics[width=0.7\textwidth]{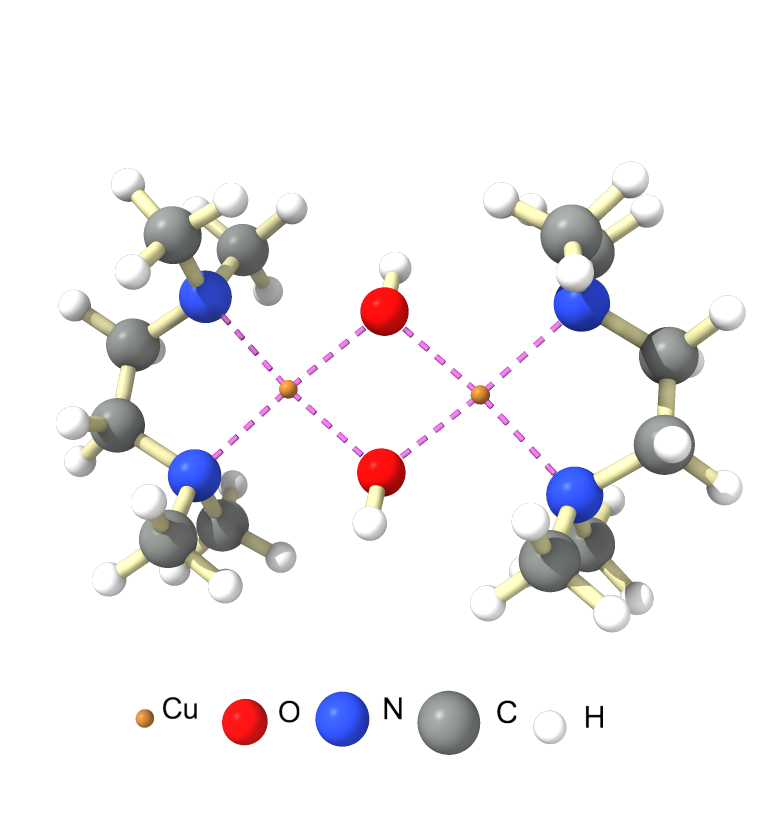}
\end{figure}

As a realistic test case, we finally examine the bis($\mu$-hydroxo) Cu(II)
complex (Fig.~\ref{fig:cu2}). This complex carries a total charge of 2 and is
treated in two spin configurations: a triplet HS state (\(S=1\)) and a LS
singlet state (\(S=0\)) modeled via a BS solution with antiferromagnetically
aligned local moments on the two Cu(II) centers. The two copper atoms are
separated by approximately 2.97~\AA, and the magnetic interaction is mediated by
bridging ligands (oxygen donors) in a non-collinear geometry, giving rise to a
more delocalized and angle-dependent superexchange pathway than in the linear
\ce{H2He} model. Due to the partially filled Cu \(d\)-manifold and the
multireference character of the singlet, the BS state can exhibit substantial
spin contamination.  We therefore apply the constraint to enforce the target
spin character of the BS state (\(\langle\hat{S}^2\rangle\)=1) and assess its
impact on the computed exchange couplings.
Table~\ref{cu2table} shows the calculated magnetic exchange couplings  for the
bis($\mu$-hydroxo) Cu(II) complex. The values of \(\langle \hat{S}^2 \rangle\)
obtained for the HS state are very close to the ideal spin-pure value,
indicating minimal spin contamination. In contrast, the BS solutions exhibit
substantial deviations from the ideal value of \(\langle \hat{S}^2 \rangle =
1\), with the exception of the two hybrid functionals employed, PBEh and B3LYP,
for which the BS spin expectation values are already close to 1. For the BLYP
and PBE functionals, the application of the spin-squared constraint leads to a
systematic reduction in the magnitude of the exchange coupling constants,
consistent with the trends observed for the simpler \ce{H2He} and \ce{H3He3}
systems. This behavior reflects the energetic penalty imposed by the constraint
on the spin-contaminated BS state, which increases the BS energy relative to the
HS reference and consequently lowers the resulting exchange coupling.
\begin{table}[ht]
	\begin{ruledtabular}
		\caption{\(J_N\), \(J_R\), and \(J_Y\) (in cm\(^{-1}\)) for the
			bis($\mu$-hydroxo) Cu(II) complex.}
		\label{cu2table}
		\begin{tabular}{lccccccr}
			Functional & \(\langle\hat{S}^2\rangle_\text{HS}\) & \(\langle\hat{S}^2\rangle_\text{BS}\) & \(J_N\) & \(J_R\) & \(J_Y\) & \(J_c\) & Experimental\footnote{Taken from Ref.~\citenum{singh2021}} \\
			\hline
			BLYP       & 2.00293                               & 0.80620                               & -743    & -371    & -621    & -342    & -180                                                       \\
			PBE        & 2.00299                               & 0.81101                               & -729    & -364    & -612    & -342                                                                 \\
			PBEh       & 2.00646                               & 0.98751                               & -167    & -83     & -164    & -724                                                                 \\
			B3LYP      & 2.00559                               & 0.97732                               & -225    & -112    & -219    & -638                                                                 \\
		\end{tabular}
	\end{ruledtabular} \end{table}
In contrast to the local and semi-local functionals, the hybrid functionals PBEh
and B3LYP exhibit a qualitatively different response to the spin-squared
constraint. In this case, the constrained \(J_c\)  increases in magnitude. This
indicates that, for hybrid functionals, the application of the constraint
modifies the energetic balance between the HS and BS states in a manner distinct
from that observed for semi-local approximations. This behavior can be
rationalized by considering the electronic structure of the BS state obtained
with hybrid functionals. In general, the spin-squared constraint has a larger
energetic impact on the BS solution than on the HS reference, increasing the BS
energy and thereby reducing the HS–BS energy gap and the resulting exchange
coupling constant. However, hybrid functionals partially correct
self-interaction errors and promote stronger localization of the spin density,
yielding BS solutions that are intrinsically closer to the symmetry-broken
limit. As reflected by the \(\langle \hat{S}^2 \rangle\) values reported in the
third column of Table~\ref{cu2table}, the unconstrained BS states obtained with
hybrid functionals are already close to the ideal value of 1. Consequently, the
additional constraint has a comparatively small effect on the BS energy than on
the HS energy.  The net effect is an increase in the HS–BS energy gap, which
leads to larger values of the exchange coupling constant. This mechanism is
illustrated by the marked difference between the unconstrained BS spin
expectation values obtained with BLYP (\(\langle \hat{S}^2 \rangle \approx
0.80\)) and those obtained with B3LYP (\(\langle \hat{S}^2 \rangle \approx
0.97\)), the latter being much closer to the ideal BS value. This trend was not
observed in the \ce{H2He} and \ce{H3He3} molecules.

Overall, the bis($\mu$-hydroxo) Cu(II) results demonstrate that the spin-squared
constraint is an effective tool for correcting the systematic overestimation of
exchange interactions in local and semi-local density functionals. For hybrid
functionals, however, the constraint must be applied with  care, as the improved
treatment of self-interaction and enhanced spin localization fundamentally alter
the balance between spin contamination, energetic penalties, and magnetic
coupling.

\section{Conclusions}\label{sec:conclusions}

In this work, we formulated a constrained DFT approach within the GKS framework
that enables explicit control of \(\langle \hat{S}^2 \rangle\) through a
Lagrange multiplier. Rather than eliminating spin contamination in BS solutions,
the method enforces physically motivated spin mixing consistent with BS-based
exchange-coupling models. For example, the target values \(\langle \hat{S}^2
\rangle = 1\) (singlet--triplet) and \(\langle \hat{S}^2 \rangle = 1.75\)
(doublet--quartet) are recovered by construction. We also derived analytical
gradients of \(\langle \hat{S}^2 \rangle\) with respect to spin-resolved density
matrices, enabling an efficient SCF implementation. These expressions are
general for arbitrary spin manifolds and do not rely on a specific
exchange-correlation functional, providing a practical route for constrained BS
calculations across different chemical systems.

Applied to \ce{H2He}, \ce{H3He3}, and the bis($\mu$-hydroxo) Cu(II) complex
exchange couplings, the constraint reduces functional-dependent variability
in calculated \(J\) couplings and stabilizes the spin character of BS solutions.
The constrained \(J\) values are generally smaller in magnitude than unconstrained
ones, particularly for semi-local functionals, where spin contamination is more
pronounced.
An exception is the bis($\mu$-hydroxo) Cu(II) complex with hybrid functionals
(PBEh, B3LYP), where unconstrained BS solutions are already close to \(\langle
\hat{S}^2 \rangle \approx 1\), and the constraint can increase \(|J|\) through a
stronger impact on the HS reference.

\section{Acknowledgments}\label{sec:Acknowledgment}

This work was supported by the U.S. Department of Energy, Office of Science, Office of Basic Energy Sciences, as part of the Computational Chemical Sciences Program under Award No. DE-SC0005027.

\appendix
\section{The effect of \(\lambda\)}\label{appendixa}

\begin{figure}[ht]
	\centering
	\begin{subfigure}[b]{0.49\textwidth}
		\centering
		\includegraphics[width=\textwidth]{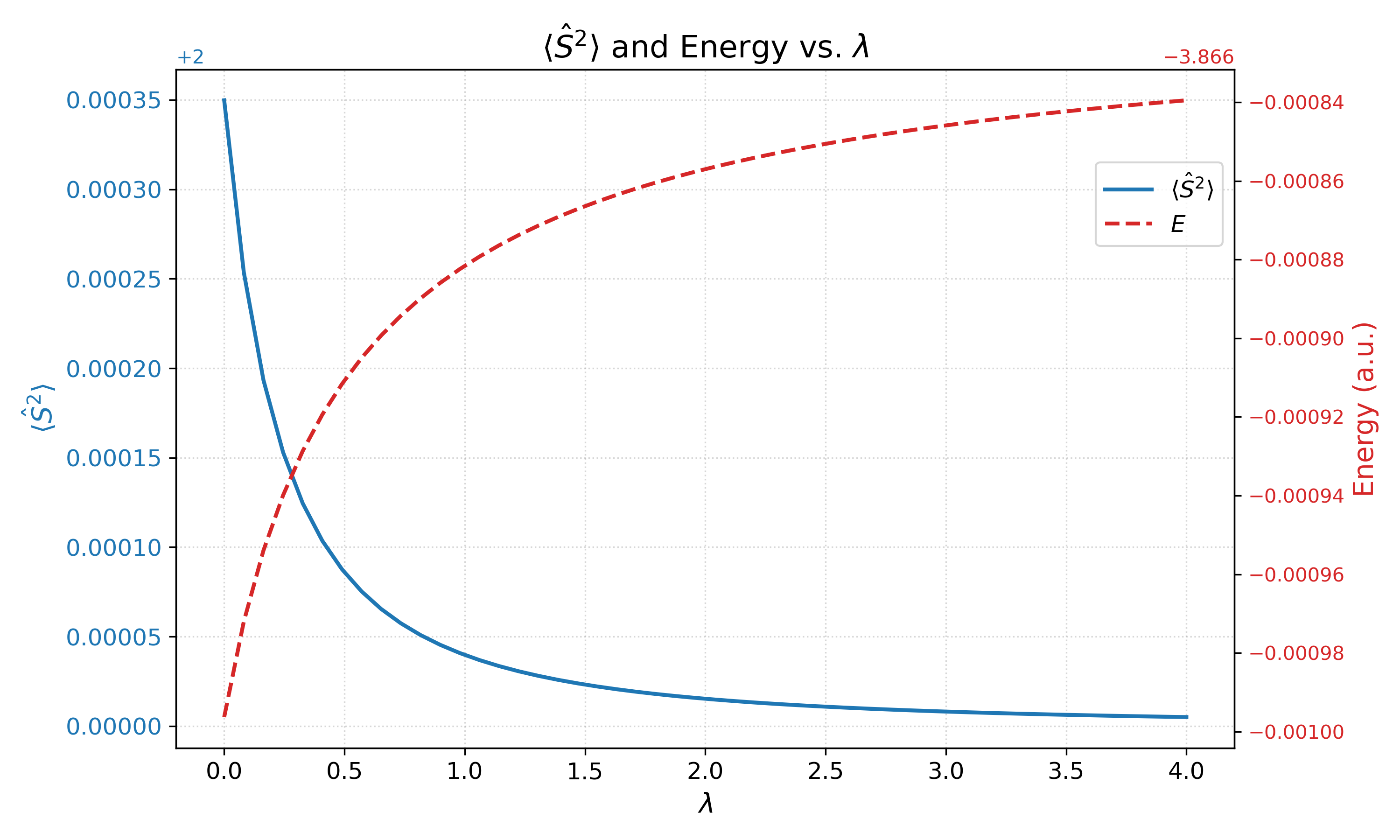}
		\caption{}
		\label{fig:fmplot}
	\end{subfigure}
	\hfill
	\begin{subfigure}[b]{0.49\textwidth}
		\centering
		\includegraphics[width=\textwidth]{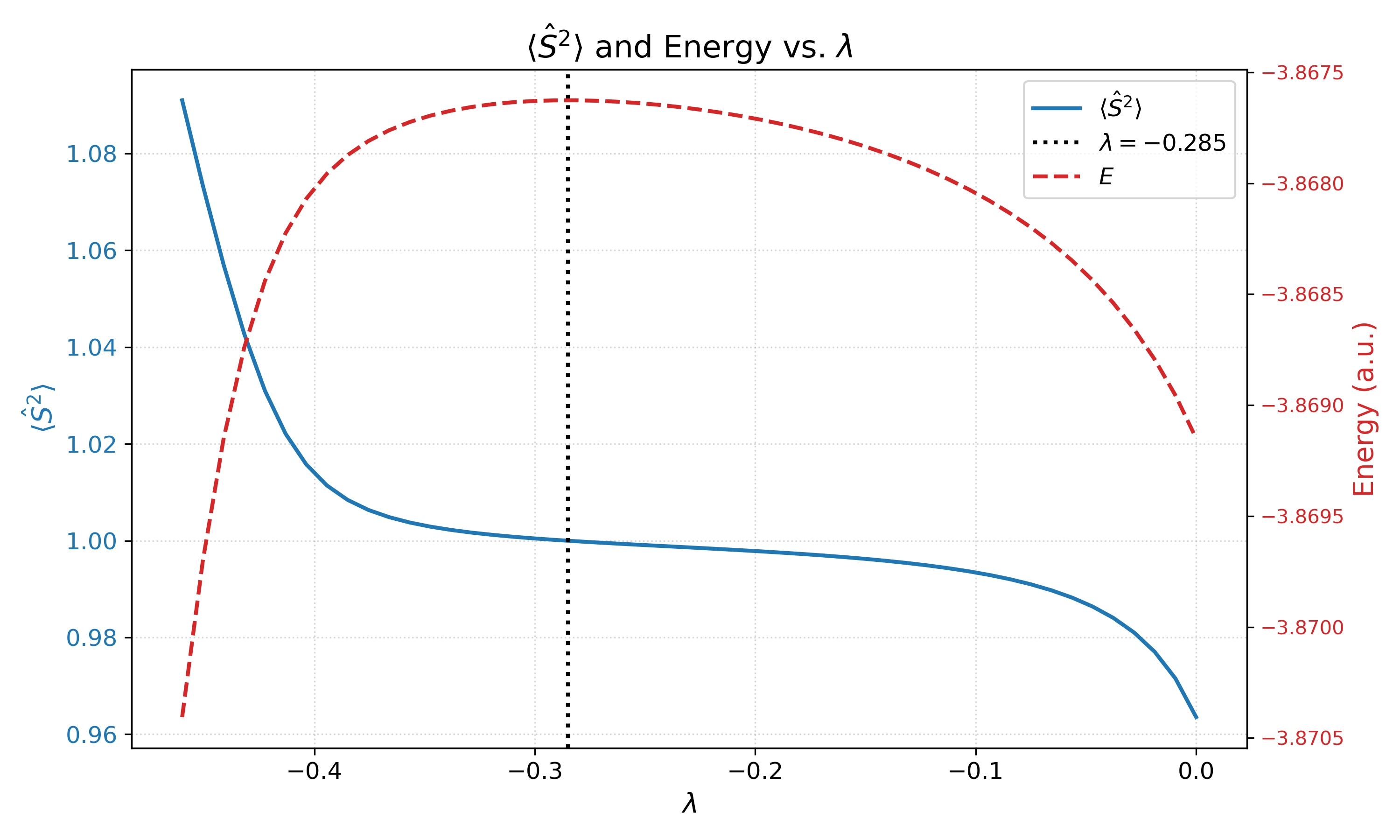}
		\caption{}
		\label{fig:afplot}
	\end{subfigure}

	\vspace{0.5cm}
	\begin{subfigure}[b]{0.49\textwidth}
		\centering
		\includegraphics[width=\textwidth]{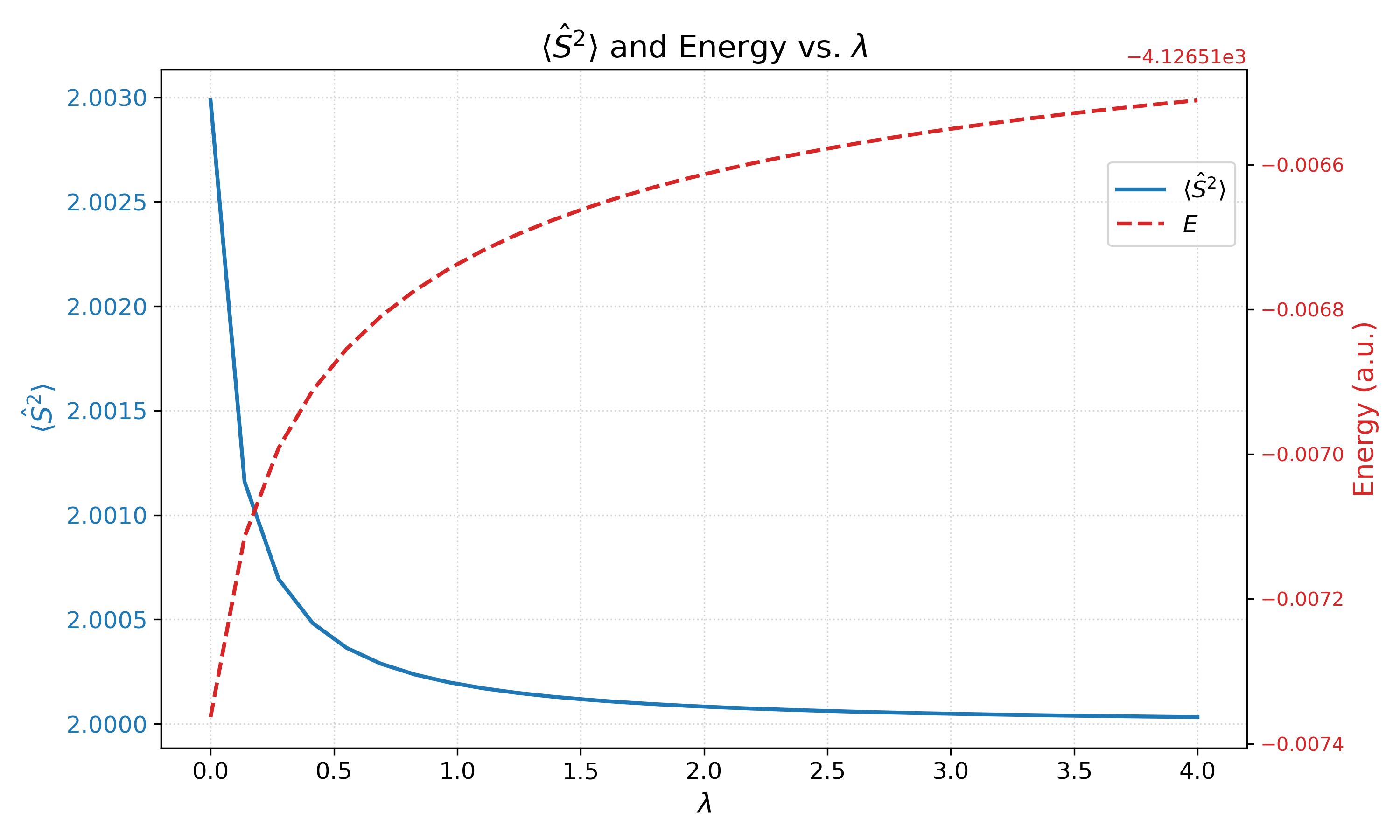}
		\caption{}
		\label{fig:cu_fmplot}
	\end{subfigure}
	\hfill
	\begin{subfigure}[b]{0.49\textwidth}
		\centering
		\includegraphics[width=\textwidth]{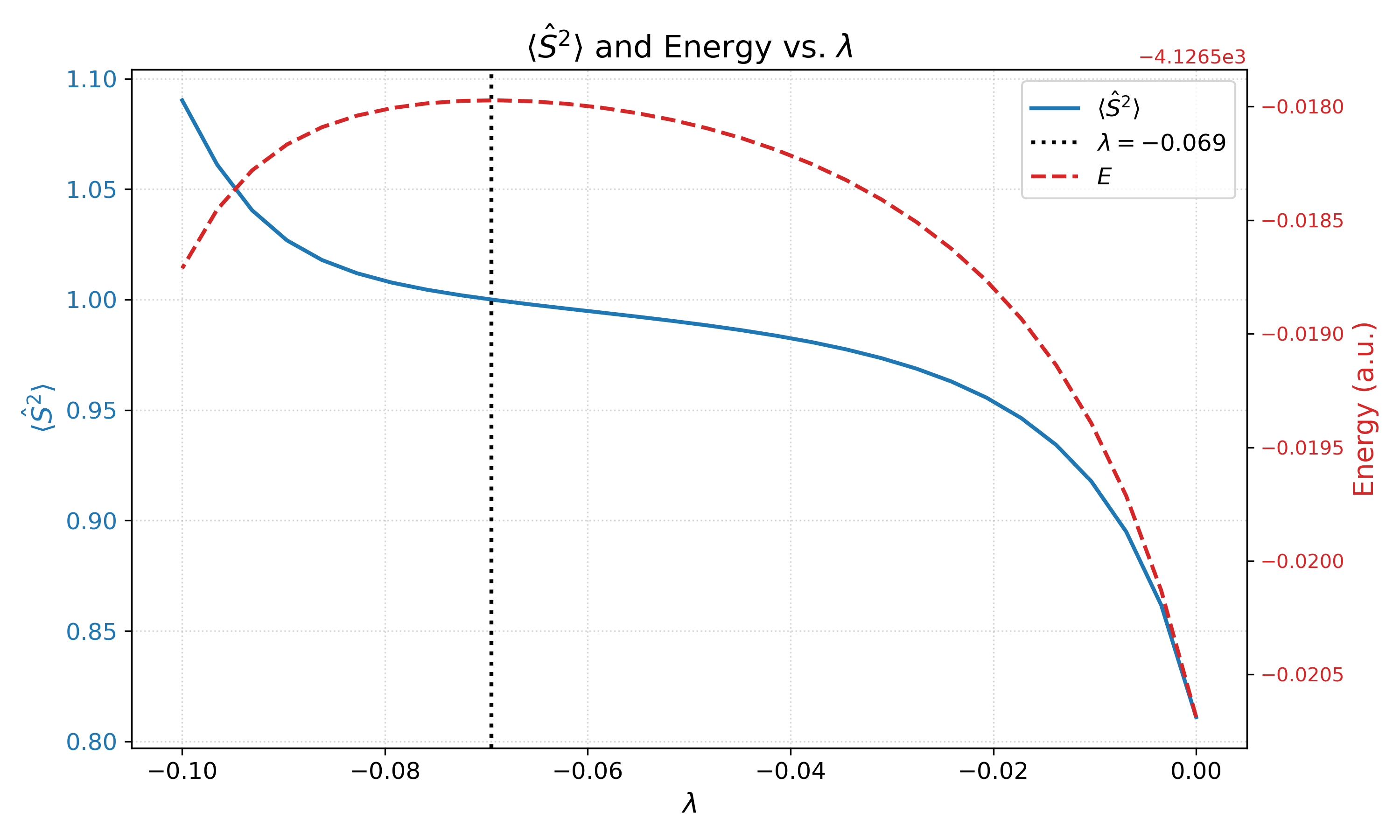}
		\caption{}
		\label{fig:cu_afplot}
	\end{subfigure}

	\caption{Total energy and \(\langle \hat{S}^2 \rangle\) as functions
		of \(\lambda\) for HS (left) and BS (right) states in \ce{H2He} (top)
		and the bis($\mu$-hydroxo) Cu(II) complex (bottom). HS curves are
		monotonic, while BS curves
		show a stationary point.}
	\label{fig:plot}
\end{figure}

To clarify the role of the Lagrange multiplier in the spin-squared constrained
formalism, we examine how both the total energy and \(\langle \hat{S}^2
\rangle\) change as \(\lambda\) is varied. Figure~\ref{fig:plot} summarizes this
dependence for HS and BS solutions in \ce{H2He} and the bis($\mu$-hydroxo)
Cu(II) complex, allowing a direct comparison of the constraint’s impact on the
two spin manifolds. As shown below, the response is qualitatively different in
the HS and BS cases, which has important consequences for how \(\lambda\) should
be chosen and optimized in practical calculations.
For the HS state in the \ce{H2He} molecule and the bis($\mu$-hydroxo) Cu(II)
complex (Figs.~\ref{fig:fmplot} and \ref{fig:cu_fmplot} respectively), both the
total energy and \(\langle \hat{S}^2 \rangle\) exhibit a smooth and monotonic
dependence on the Lagrange multiplier \(\lambda\). As \(\lambda \rightarrow
\infty\), both quantities approach well-defined asymptotic limits, reflecting
the progressive elimination of residual spin contamination in the HS solution.
This behavior indicates that an explicit optimization of the constrained
Lagrangian with respect to \(\lambda\) is not required for the HS state:
selecting a sufficiently large multiplier is sufficient to recover a spin-pure
reference. Although spin contamination in HS states is typically small, the
calculation of \(J\) is highly sensitive to even minor deviations in \(\langle
\hat{S}^2 \rangle\), making HS spin decontamination advisable.

Insight into this behavior can be obtained by examining Eq.~\ref{eq:s2}. The
spin-contamination contribution vanishes when the condition
\begin{equation}
	N_\beta = \sum_{\mu\nu\kappa\lambda}
	P^\alpha_{\mu\nu} O_{\nu\kappa} P^\beta_{\kappa\lambda} O_{\lambda\mu}
	\label{eq:nbetatrace}
\end{equation}
is satisfied. The trace on the right-hand side can be rewritten as
\begin{equation}
	\sum_{i=1}^{N_\alpha} \sum_{j=1}^{N_\beta}
	\left| \braket{\phi^\alpha_i}{\phi^\beta_j} \right|^2 ,
\end{equation}
which might suggest that Eq.~\ref{eq:nbetatrace} holds only when the occupied
\(\alpha\) and \(\beta\) orbitals are identical. This interpretation would imply
that complete orbital pairing is required to eliminate spin contamination.

A closely related constraint was previously introduced by Andrews et
al.,~\cite{andrews1991spin} within the Hartree–Fock framework, where the
restriction was imposed directly on the trace term in Eq.~\ref{eq:s2}. They
demonstrated that, in the limit \(\lambda \rightarrow \infty\), restricted
open-shell Hartree–Fock solutions are obtained. While this result is consistent
with the interpretation above, the present analysis shows that
Eq.~\ref{eq:nbetatrace} does not require such a strong condition. In particular,
the equality holds whenever the occupied \(\beta\) manifold can be expressed as
linear combinations of the occupied \(\alpha\) orbitals, without requiring
one-to-one orbital pairing. A formal proof of this result is provided in
Appendix~\ref{appendixb}.

In contrast, the BS solution for the \ce{H2He} molecule (Fig.~\ref{fig:afplot})
and for the bis($\mu$-hydroxo) Cu(II) complex (Fig.~\ref{fig:cu_afplot})
displays a qualitatively different behavior. Both the energy and \(\langle
\hat{S}^2 \rangle\) exhibit a stationary point as functions of \(\lambda\),
making an explicit optimization with respect to the Lagrange multiplier
necessary.
Performing the optimization of the Lagrangian with respect to \(\lambda\) in
\ce{H2He} yields a stationary point at \(\lambda \approx -0.28\), indicated by
the vertical dashed line in Fig.~\ref{fig:plot}. In the case of the
bis($\mu$-hydroxo) Cu(II) complex, the maximum point is at \(\lambda \approx
-0.07\). These stationary points correspond to a maximum of the constrained
energy, reflecting the fact that the constraint penalizes deviations from the
target spin value. At this optimal \(\lambda\), the BS solution satisfies the
imposed condition \(\langle \hat{S}^2 \rangle = 1\) exactly, providing a
consistent and well-defined low-spin reference for the evaluation of \(J\).

\section{Conditions for \(\mathrm{Tr}(\mathbf{P}^\alpha \mathbf{P}^\beta) = N_\beta\)}\label{appendixb}

Consider Eq.~\ref{eq:s2} and, for clarity, let us work on an orthonormal
basis. In this case, the overlap matrix reduces to the identity and the
expectation value of $\hat{S}^2$ can be written as
\begin{align}
	\langle \hat{S}^2 \rangle & =
	S_z(S_z + 1) + N_\beta - \sum_{\mu\nu} P^\alpha_{\mu\nu} P^\beta_{\nu\mu}                                   \\
	                          & = S_z(S_z + 1) + N_\beta - \mathrm{Tr}(\mathbf{P}^\alpha \mathbf{P}^\beta) \; .
\end{align}
Since the trace is invariant under unitary transformations, the following
derivation is fully equivalent to the general expression in Eq.~\ref{eq:s2}
written in a nonorthogonal atomic-orbital basis.

The spin-resolved density matrices can be expressed as projectors onto the
subspaces spanned by the occupied spatial orbitals of each spin,
\begin{equation}
	\mathbf{P}^\alpha = \sum_{i=1}^{N_\alpha} \ket{\phi^\alpha_i}\bra{\phi^\alpha_i},
	\label{eq:projectora}
\end{equation}
\begin{equation}
	\mathbf{P}^\beta = \sum_{j=1}^{N_\beta} \ket{\phi^\beta_j}\bra{\phi^\beta_j},
	\label{eq:projectorb}
\end{equation}
where $\{\phi_i^\alpha\}$ and $\{\phi_j^\beta\}$ denote orthonormal sets of
occupied spatial orbitals.

Using these definitions, it is straightforward to show that
\begin{equation}
	\mathrm{Tr}(\mathbf{P}^\alpha \mathbf{P}^\beta)
	= \sum_{i=1}^{N_\alpha} \sum_{j=1}^{N_\beta}
	\left| \braket{\phi^\alpha_i}{\phi^\beta_j} \right|^2,
	\label{eq:projectortrace}
\end{equation}
which depends only on the mutual overlaps between the occupied $\alpha$ and
$\beta$ spatial orbitals.

Introducing the overlap matrix between the two orbital sets,
\begin{equation}
	O_{ij} = \braket{\phi^\alpha_i}{\phi^\beta_j},
\end{equation}
Eq.~\eqref{eq:projectortrace} can be written in compact form as
\begin{equation}
	\mathrm{Tr}(\mathbf{P}^\alpha \mathbf{P}^\beta)
	= \mathrm{Tr}(\mathbf{O}^\dagger \mathbf{O})\; .
	\label{eq:traceinoperators}
\end{equation}
We now assume that the occupied $\beta$ orbitals are contained in the subspace
spanned by the occupied $\alpha$ orbitals, such that each $\beta$ orbital can be
expressed as a linear combination of the $\alpha$ orbitals,
\begin{equation}
	\ket{\phi_j^\beta} = \sum_{i=1}^{N_\alpha} C_{ij} \ket{\phi_i^\alpha}.
	\label{eq:linearcomb}
\end{equation}
This condition reflects a geometric inclusion of subspaces and does not imply
equality or pairing of individual $\alpha$ and $\beta$ orbitals.

Using the orthonormality of the $\alpha$ orbitals, the orthonormality of the
$\beta$ orbitals implies
\begin{align}
	\delta_{jj'} & = \braket{\phi_j^\beta}{\phi_{j'}^\beta}                                 \\
	             & = \sum_{i,i'} C_{ij}^* C_{i'j'} \braket{\phi_i^\alpha}{\phi_{i'}^\alpha} \\
	             & = \sum_{i} C_{ij}^* C_{ij'},
\end{align}
which, in matrix form, yields
\begin{equation}
	\mathbf{C}^\dagger \mathbf{C} = \mathbf{1}_{N_\beta}.
\end{equation}
Substituting Eq.~\eqref{eq:linearcomb} into the definition of the overlap
matrix,
\begin{align}
	O_{ij} & = \braket{\phi^\alpha_i}{\phi^\beta_j}                               \\
	       & = \sum_{k=1}^{N_\alpha} C_{kj} \braket{\phi_i^\alpha}{\phi_k^\alpha} \\
	       & = C_{ij},
\end{align}
so that $\mathbf{O} = \mathbf{C}$. Consequently,
\begin{equation}
	\mathrm{Tr}(\mathbf{O}^\dagger \mathbf{O})
	= \mathrm{Tr}(\mathbf{C}^\dagger \mathbf{C})
	= \mathrm{Tr}(\mathbf{1}_{N_\beta})
	= N_\beta.
\end{equation}
Therefore, when the occupied $\beta$ manifold are entirely contained within the
subspace spanned by the occupied $\alpha$ orbitals, the trace term in
Eq.~\ref{eq:s2} equals the number of $\beta$ electrons. In this case, the
additional contribution to $\langle \hat{S}^2 \rangle$ vanishes and the Slater
determinant is an eigenfunction of $\hat{S}^2$, yielding the correct spin value
without spin contamination.

\bibliography{Magnetism_abbrev}

@article{Hohenberg1964,
  title = {Inhomogeneous Electron Gas},
  author = {Hohenberg, P. and Kohn, W.},
  journal = {Phys. Rev.},
  volume = {136},
  issue = {3B},
  pages = {B864--B871},
  year = {1964},
  month = nov,
  publisher = {American Physical Society},
  doi = {10.1103/PhysRev.136.B864},
}

@article{Kohn1965,
  title = {Self-Consistent Equations Including Exchange and Correlation Effects},
  author = {Kohn, W. and Sham, L. J.},
  journal = {Phys. Rev.},
  volume = {140},
  issue = {4A},
  pages = {A1133--A1138},
  year = {1965},
  month = nov,
  publisher = {American Physical Society},
  doi = {10.1103/PhysRev.140.A1133},
}

@article{Trepte2021,
    author = {Trepte, Kai and Schwalbe, Sebastian and Liebing, Simon and Schulze, Wanja T. and Kortus, Jens and Myneni, Hemanadhan and Ivanov, Aleksei V. and Lehtola, Susi},
    title = {Chemical bonding theories as guides for self-interaction corrected solutions: Multiple local minima and symmetry breaking},
    journal = {J. Chem. Phys.},
    volume = {155},
    number = {22},
    pages = {224109},
    year = {2021},
    month = dec,
    issn = {0021-9606},
    doi = {10.1063/5.0071796},
}

@article{Ruiz2005,
    author = {Ruiz, Eliseo and Alvarez, Santiago and Cano, Joan and Polo, Víctor},
    title = {About the calculation of exchange coupling constants using density-functional theory: The role of the self-interaction error},
    journal = {J. Chem. Phys.},
    volume = {123},
    number = {16},
    pages = {164110},
    year = {2005},
    month = oct,
    issn = {0021-9606},
    doi = {10.1063/1.2085171},
}

@article{Bryenton2023,
author = {Bryenton, Kyle R. and Adeleke, Adebayo A. and Dale, Stephen G. and Johnson, Erin R.},
title = {Delocalization error: The greatest outstanding challenge in density-functional theory},
 journal = {WIREs Comput. Mol. Sci.},
volume = {13},
number = {2},
pages = {e1631},
 doi = {10.1002/wcms.1631},
year = {2023}
}

@article{Mori2008,
  title = {Localization and Delocalization Errors in Density Functional Theory and Implications for Band-Gap Prediction},
  author = {Mori-S\'anchez, Paula and Cohen, Aron J. and Yang, Weitao},
  journal = {Phys. Rev. Lett.},
  volume = {100},
  issue = {14},
  pages = {146401},
  numpages = {4},
  year = {2008},
  month = apr,
  publisher = {American Physical Society},
  doi = {10.1103/PhysRevLett.100.146401},
}

@incollection{herbert2023,
  title={Density-functional theory for electronic excited states},
  author={Herbert, John M},
  booktitle={Theoretical and computational photochemistry},
  pages={69--118},
  year={2023},
  publisher={Elsevier}
}

@article{kaduk2012,
  title={Constrained density functional theory},
  author={Kaduk, Benjamin and Kowalczyk, Tim and Van Voorhis, Troy},
  journal={Chem. Rev.},
  volume={112},
  number={1},
  pages={321--370},
  year={2012},
  publisher={ACS Publications}
}

@article{wu2006,
  title={Constrained density functional theory and its application in long-range electron transfer},
  author={Wu, Qin and Van Voorhis, Troy},
  journal={J. Chem. Theory Comput.},
  volume={2},
  number={3},
  pages={765--774},
  year={2006},
  publisher={ACS Publications}
}

@article{adamo2013,
  title={The calculations of excited-state properties with Time-Dependent Density Functional Theory},
  author={Adamo, Carlo and Jacquemin, Denis},
  journal={Chem. Soc. Rev.},
  volume={42},
  number={3},
  pages={845--856},
  year={2013},
  publisher={Royal Society of Chemistry}
}

@Article{von1972local,
	author = {von Barth, Ulf and Hedin, Lars},
	journal = {J. Phys. C: Solid State Phys.},
	number = {13},
	pages = {1629},
	publisher = {IOP Publishing},
	title = {{A local exchange-correlation potential for the spin polarized case. I}},
	volume = {5},
	year = {1972}}

@article{perdew1981,
  title = {Self-interaction correction to density-functional approximations for many-electron systems},
  author = {Perdew, J. P. and Zunger, Alex},
  journal = {Phys. Rev. B},
  volume = {23},
  issue = {10},
  pages = {5048--5079},
  year = {1981},
  month = may,
  publisher = {American Physical Society},
  doi = {10.1103/PhysRevB.23.5048},
}

@book{chattaraj2009chemical,
	author = {Chattaraj, Pratim Kumar},
	editor = {Vargas, Rubicelia and Galv{\'{a}}n, Marcelo},
	publisher = {CRC press},
	title = {{Spin-Polarized Density Functional Theory: Chemical Reactivity . Chemical reactivity theory: a density functional view}},
	year = {2009}}

@article{pyscf2020,
    author = {Sun, Qiming and Zhang, Xing and Banerjee, Samragni and Bao, Peng and Barbry, Marc and Blunt, Nick S. and Bogdanov, Nikolay A. and Booth, George H. and Chen, Jia and Cui, Zhi-Hao and Eriksen, Janus J. and Gao, Yang and Guo, Sheng and Hermann, Jan and Hermes, Matthew R. and Koh, Kevin and Koval, Peter and Lehtola, Susi and Li, Zhendong and Liu, Junzi and Mardirossian, Narbe and McClain, James D. and Motta, Mario and Mussard, Bastien and Pham, Hung Q. and Pulkin, Artem and Purwanto, Wirawan and Robinson, Paul J. and Ronca, Enrico and Sayfutyarova, Elvira R. and Scheurer, Maximilian and Schurkus, Henry F. and Smith, James E. T. and Sun, Chong and Sun, Shi-Ning and Upadhyay, Shiv and Wagner, Lucas K. and Wang, Xiao and White, Alec and Whitfield, James Daniel and Williamson, Mark J. and Wouters, Sebastian and Yang, Jun and Yu, Jason M. and Zhu, Tianyu and Berkelbach, Timothy C. and Sharma, Sandeep and Sokolov, Alexander Yu. and Chan, Garnet Kin-Lic},
    title = {Recent developments in the PySCF program package},
    journal = {J. Chem. Phys.},
    volume = {153},
    number = {2},
    pages = {024109},
    year = {2020},
    month = jul,
    issn = {0021-9606},
    doi = {10.1063/5.0006074},
}

@article{LEHTOLA20181,
title = {Recent developments in libxc — A comprehensive library of functionals for density functional theory},
journal = {SoftwareX},
volume = {7},
pages = {1-5},
year = {2018},
issn = {2352-7110},
doi = {10.1016/j.softx.2017.11.002},
author = {Susi Lehtola and Conrad Steigemann and Micael J.T. Oliveira and Miguel A.L. Marques}
}

@article{dederichs1984ground,
  title={Ground states of constrained systems: application to cerium impurities},
  author={Dederichs, PH and Bl{\"u}gel, S and Zeller, R and Akai, H},
  journal={Phys. Rev. Lett.},
  volume={53},
  number={26},
  pages={2512},
  year={1984},
  publisher={APS}
}

@article{stein2009reliable,
  title={Reliable prediction of charge transfer excitations in molecular complexes using time-dependent density functional theory},
  author={Stein, Tamar and Kronik, Leeor and Baer, Roi},
  journal={J. Am. Chem. Soc.},
  volume={131},
  number={8},
  pages={2818--2820},
  year={2009},
  publisher={ACS Publications}
}

@article{gould2021ensemble,
  title={Ensemble generalized Kohn--Sham theory: The good, the bad, and the ugly},
  author={Gould, Tim and Kronik, Leeor},
  journal={J. Chem. Phys.},
  volume={154},
  number={9},
  year={2021},
  publisher={AIP Publishing}
}

@article{sim2022improving,
  title={Improving results by improving densities: Density-corrected density functional theory},
  author={Sim, Eunji and Song, Suhwan and Vuckovic, Stefan and Burke, Kieron},
  journal={J. Am. Chem. Soc.},
  volume={144},
  number={15},
  pages={6625--6639},
  year={2022},
  publisher={ACS Publications}
}

@article{gaggioli2019beyond,
  title={Beyond density functional theory: the multiconfigurational approach to model heterogeneous catalysis},
  author={Gaggioli, Carlo Alberto and Stoneburner, Samuel J and Cramer, Christopher J and Gagliardi, Laura},
  journal={ACS Catal.},
  volume={9},
  number={9},
  pages={8481--8502},
  year={2019},
  publisher={ACS Publications}
}

@article{wu2006direct,
  title={Direct calculation of electron transfer parameters through constrained density functional theory},
  author={Wu, Qin and Van Voorhis, Troy},
  journal={J. Phys. Chem. A},
  volume={110},
  number={29},
  pages={9212--9218},
  year={2006},
  publisher={ACS Publications}
}

@article{fonseca1998self,
  title={Self-consistent electronic structure, coulomb interaction, and spin effects in self-assembled strained InAs--GaAs quantum dot structures},
  author={Fonseca, LRCM and Jimenez, JL and Leburton, JP and Martin, Richard M},
  journal={Physica E},
  volume={2},
  number={1-4},
  pages={743--747},
  year={1998},
  publisher={Elsevier}
}

@article{eriksson2005many,
  title={Many-body projector orbitals for electronic structure theory of strongly correlated electrons},
  author={Eriksson, Olle and Wills, John M and Colarieti-Tosti, Massimiliano and Lebegue, Sebastien and Grechnev, Alexei},
  journal={Int. J. Quantum Chem.},
  volume={105},
  number={2},
  pages={160--165},
  year={2005},
  publisher={Wiley Online Library}
}

@inproceedings{hourahine2010dftb+,
  title={{DFTB+ and lanthanides}},
  author={Hourahine, B and Aradi, B and Frauenheim, T},
  booktitle={J. Phys.: Conf. Ser.},
  volume={242},
  number={1},
  pages={012005},
  year={2010},
  organization={IOP Publishing}
}

@article{phillips2011magnetic,
  title={Magnetic exchange couplings from constrained density functional theory: An efficient approach utilizing analytic derivatives},
  author={Phillips, Jordan J and Peralta, Juan E},
  journal={J. Chem. Phys.},
  volume={135},
  number={18},
  year={2011},
  publisher={AIP Publishing}
}

@article{ni2025visible,
  title={Visible light enhanced thermocatalytic reverse water gas shift reaction via localized surface plasmon resonance of copper nanoparticles},
  author={Ni, Wenkang and Zhang, Xiaoyan and Yue, Xuanyu and Zhang, Zizhong and Zhang, Yongfan and Wang, Ke and Dai, Wenxin and Fu, Xianzhi},
  journal={Sep. Purif. Technol.},
  volume={361},
  pages={131514},
  year={2025},
  publisher={Elsevier}
}

@article{ruiz1999broken,
  title={Broken symmetry approach to calculation of exchange coupling constants for homobinuclear and heterobinuclear transition metal complexes},
  author={Ruiz, Eliseo and Cano, Joan and Alvarez, Santiago and Alemany, Pere},
  journal={J. Comput. Chem.},
  volume={20},
  number={13},
  pages={1391--1400},
  year={1999},
  publisher={Wiley Online Library}
}

@article{ferre2015spin,
  title={Spin decontamination of broken-symmetry density functional theory calculations: deeper insight and new formulations},
  author={Ferr{\'e}, Nicolas and Guih{\'e}ry, Nathalie and Malrieu, Jean-Paul},
  journal={Phys. Chem. Chem. Phys.},
  volume={17},
  number={22},
  pages={14375--14382},
  year={2015},
  publisher={Royal Society of Chemistry}
}

@article{david2020consistent,
  title={Consistent spin decontamination of broken-symmetry calculations of diradicals},
  author={David, Gr{\'e}goire and Trinquier, Georges and Malrieu, Jean-Paul},
  journal={J. Chem. Phys.},
  volume={153},
  number={19},
  year={2020},
  publisher={AIP Publishing}
}

@article{soda2000ab,
  title={{Ab initio computations of effective exchange integrals for H--H,
  H--He--H and Mn2O2 complex: comparison of broken-symmetry approaches}},
  author={Soda, T and Kitagawa, Y and Onishi, T and Takano, Y and Shigeta, Y and Nagao, H and Yoshioka, Y and Yamaguchi, K},
  journal={Chem. Phys. Lett.},
  volume={319},
  number={3-4},
  pages={223--230},
  year={2000},
  publisher={Elsevier}
}

@article{singh2021,
  title={Magnetic exchange coupling in Cu dimers studied with modern multireference methods and broken-symmetry coupled cluster theory},
  author={Singh, Gurjot and Gamboa, Stefani and Orio, Maylis and Pantazis, Dimitrios A and Roemelt, Michael},
  journal={Theor. Chem. Acc.},
  volume={140},
  pages={1--15},
  year={2021},
  publisher={Springer}
}

@article{noodleman1981valence,
  title={Valence bond description of antiferromagnetic coupling in transition metal dimers},
  author={Noodleman, Louis},
  journal={J. Chem. Phys.},
  volume={74},
  number={10},
  pages={5737--5743},
  year={1981},
  publisher={American Institute of Physics}
}

@article{ovchinnikov1996simple,
  title={Simple spin correction of unrestricted density-functional calculation},
  author={Ovchinnikov, Alexander A and Labanowski, Jan K},
  journal={Phys. Rev. A},
  volume={53},
  number={6},
  pages={3946},
  year={1996},
  publisher={APS}
}

@article{andrews1991spin,
  title={Spin contamination in single-determinant wavefunctions},
  author={Andrews, Jamie S and Jayatilaka, Dylan and Bone, Richard GA and Handy, Nicholas C and Amos, Roger D},
  journal={Chem. Phys. Lett.},
  volume={183},
  number={5},
  pages={423--431},
  year={1991},
  publisher={Elsevier}
}

@article{tsuchimochi2011constrained,
  title={Constrained active space unrestricted mean-field methods for controlling spin-contamination},
  author={Tsuchimochi, Takashi and Scuseria, Gustavo E},
  journal={J. Chem. Phys.},
  volume={134},
  number={6},
  year={2011},
  publisher={AIP Publishing}
}

@article{lehtola2020overview,
  title={An overview of self-consistent field calculations within finite basis sets},
  author={Lehtola, Susi and Blockhuys, Frank and Van Alsenoy, Christian},
  journal={Molecules},
  volume={25},
  number={5},
  pages={1218},
  year={2020},
  publisher={MDPI}
}

@article{schmidt2008controlling,
  title={Controlling spin contamination using constrained density functional theory},
  author={Schmidt, JR and Shenvi, Neil and Tully, John C},
  journal={J. Chem. Phys.},
  volume={129},
  number={11},
  year={2008},
  publisher={AIP Publishing}
}

@article{perdew2025scan,
  title={{SCAN meta-GGA, strong correlation, symmetry breaking, self-interaction
  correction, and semi-classical limit in density functional theory: Hidden
  connections and beneficial synergies?}}, author={Perdew, John P}, journal={APL
  Comput. Phys.}, 
  volume={1}, 
  number={1}, 
  year={2025}, 
  publisher={AIP Publishing} 
}

@article{mori2006many,
  title={Many-electron self-interaction error in approximate density functionals},
  author={Mori-S{\'a}nchez, Paula and Cohen, Aron J and Yang, Weitao},
  journal={J. Chem. Phys.},
  volume={125},
  number={20},
  year={2006},
  publisher={AIP Publishing}
}

@article{perdew1996generalized,
  title={Generalized gradient approximation made simple},
  author={Perdew, John P and Burke, Kieron and Ernzerhof, Matthias},
  journal={Phys. Rev. Lett.},
  volume={77},
  number={18},
  pages={3865},
  year={1996},
  publisher={APS}
}

@article{becke1988density,
  title={Density-functional exchange-energy approximation with correct asymptotic behavior},
  author={Becke, Axel D},
  journal={Phys. Rev. A},
  volume={38},
  number={6},
  pages={3098},
  year={1988},
  publisher={APS}
}

@article{adamo1999toward,
  title={{Toward reliable density functional methods without adjustable
  parameters: The PBE0 model}},
  author={Adamo, Carlo and Barone, Vincenzo},
  journal={J. Chem. Phys.},
  volume={110},
  number={13},
  pages={6158--6170},
  year={1999},
  publisher={American Institute of Physics}
}

@article{becke1993density,
  title={{Density-functional thermochemistry. III. The role of exact exchange}},
  author={Becke, Axel D},
  journal={J. Chem. Phys.},
  volume={98},
  number={7},
  pages={5648--5652},
  year={1993},
  publisher={American Institute of Physics}
}

@article{sun2015strongly,
  title={Strongly constrained and appropriately normed semilocal density functional},
  author={Sun, Jianwei and Ruzsinszky, Adrienn and Perdew, John P},
  journal={Phys. Rev. Lett.},
  volume={115},
  number={3},
  pages={036402},
  year={2015},
  publisher={APS}
}

@book{nocedal2006numerical,
  title={Numerical optimization},
  author={Nocedal, Jorge and Wright, Stephen J},
  year={2006},
  publisher={Springer},
  note = {See Section 6.1 (BFGS)}
}

@article{krishnan1980self,
  title={Self-consistent molecular orbital methods. XX. A basis set for correlated wave functions},
  author={Krishnan, RBJS and Binkley, J Stephen and Seeger, Rolf and Pople, John A},
  journal={J. Chem. Phys.},
  volume={72},
  number={1},
  pages={650--654},
  year={1980},
  publisher={American Institute of Physics}
}

@article{rudra2006accurate,
	title={Accurate magnetic exchange couplings in transition-metal complexes from constrained density-functional theory},
	author={Rudra, Indranil and Wu, Qin and Van Voorhis, Troy},
	journal={J. Chem. Phys.},
	volume={124},
	number={2},
	year={2006},
	publisher={AIP Publishing}
}

@article{weigend2005,
	title={Balanced basis sets of split valence, triple zeta valence and quadruple zeta valence quality for H to Rn: Design and assessment of accuracy},
	author={Weigend, Florian and Ahlrichs, Reinhart},
	journal={Phys. Chem. Chem. Phys.},
	volume={7},
	pages={3297--3305},
	year={2005},
	doi={10.1039/B508541A}
}

@article{postnikov2006density,
  title={Density functional studies of molecular magnets},
  author={Postnikov, Andrei V and Kortus, Jens and Pederson, Mark R},
  journal={phys. Status solidi (b)},
  volume={243},
  number={11},
  pages={2533--2572},
  year={2006},
  publisher={Wiley Online Library}
}

@article{caballol1997remarks,
  title={Remarks on the proper use of the broken symmetry approach to magnetic coupling},
  author={Caballol, R and Castell, O and Illas, F and {de P. R. Moreira}, I and Malrieu, JP},
  journal={J. Phys. Chem. A},
  volume={101},
  number={42},
  pages={7860--7866},
  year={1997},
  publisher={ACS Publications}
}

@article{rivero2008reliability,
  title={Reliability of range-separated hybrid functionals for describing magnetic coupling in molecular systems},
  author={Rivero, Pablo  and  de {P.R.} Moreira,  Ib{\'e}rio   and Illas, Francesc and Scuseria, Gustavo E},
  journal={J. Chem, Phys.},
  volume={129},
  number={18},
  year={2008},
  publisher={AIP Publishing}
}

@article{fitzhugh2023comparative,
  title={Comparative density functional theory study of magnetic exchange couplings in dinuclear transition-metal complexes},
  author={Fitzhugh, Henry C and Furness, James W and Pederson, Mark R and Peralta, Juan E and Sun, Jianwei},
  journal={J. Chem. Theory Comput.},
  volume={19},
  number={17},
  pages={5760--5772},
  year={2023},
  publisher={ACS Publications}
}

@article{dema2021electronic,
  title={Electronic and magnetic signatures of low-lying spin-flip excitonic states of Mn12O12-acetate},
  author={Dema, Karma and Hooshmand, Zahra and Pederson, Mark R},
  journal={Polyhedron},
  volume={206},
  pages={115332},
  year={2021},
  publisher={Elsevier}
}

@article{lowdin1964angular,
  title={Angular Momentum Wavefunctions Constructed by Projector Operators},
  author={L{\"o}wdin, Per-Olov},
  journal={Rev. Mod. Phys.},
  volume={36},
  pages={966--976},
  year={1964},
  publisher={American Physical Society}
}

@article{Moreira2006,
  title={A unified view of the theoretical description of
  magnetic coupling in molecular chemistry and solid state physics},
  author={de {P.R.} Moreira, Ib{\'e}rio and Illas, Francesc},
  journal={Phys. Chem. Chem. Phys.},
  volume={8},
  number={14},
  pages={1645--1659},
  year={2006},
  publisher={Royal Society of Chemistry}
}

\end{document}